\documentclass[11pt,a4paper]{article}

\usepackage{amsthm,amssymb,amsmath,amsfonts}
\usepackage{enumerate}
\usepackage[top=1.2in,bottom=1.2in,left=1in,right=1in]{geometry}
\usepackage[style=alphabetic,sorting=nyt,maxnames=4,maxcitenames=4,maxbibnames=4,backend=bibtex]{biblatex}
\usepackage{authblk}
\usepackage{hyperref}
\usepackage{color}
\usepackage{fancyhdr}

\theoremstyle{plain}
\newtheorem{theorem}{Theorem}[section]

\newtheorem{lemma}[theorem]{Lemma}
\newtheorem{corollary}[theorem]{Corollary}
\newtheorem{definition}[theorem]{Definition}
\newtheorem{claim}[theorem]{Claim}

\theoremstyle{definition}
\newtheorem*{remark}{Remark}

\newtheorem{example}[theorem]{Example}

\newcounter{ProtocolCounter}
\DeclareRobustCommand{\refprot}[1]{%
   \refstepcounter{ProtocolCounter}%
   \theProtocolCounter\label{#1}}
	
\let\originalparagraph\paragraph
\renewcommand{\paragraph}[2][.]{\originalparagraph{#2#1}}

\newcommand{\NN}{\mathbb{N}}

\newcommand{\RR}{\mathbb{R}}

\newcommand{\rarr}{\rightarrow}
\newcommand{\larr}{\leftarrow}

\newcommand{\dollarr}{\leftarrow}

\newcommand{\Adv}{\mathcal{A}}

\newcommand{\negl}{\mathsf{negl}}
\newcommand{\poly}{\mathsf{poly}}

\newcommand{\Enc}{\mathsf{Enc}}
\newcommand{\SGen}{\mathsf{SGen}}
\newcommand{\SEnc}{\mathsf{SEnc}}
\newcommand{\SDec}{\mathsf{SDec}}
\newcommand{\PGen}{\mathsf{PGen}}
\newcommand{\PEnc}{\mathsf{PEnc}}
\newcommand{\PDec}{\mathsf{PDec}}
\newcommand{\CCA}{\mathsf{CCA}}
\newcommand{\NM}{\mathsf{NM}}

\newcommand{\PubK}{\mathsf{PubK}}

\newcommand{\Msg}{\mathcal{M}}
\newcommand{\Ciph}{\mathcal{C}}

\newcommand{\E}{\mathop{\mathbf{E}}}
\newcommand{\eqv}[1]{\geq_{#1}}
\newcommand{\eqvrho}{\eqv{\rho}}

\newcommand{\rhoinv}{\rho^{-1}}
\newcommand{\powset}{\mathcal{P}}

\newcommand{\eps}{\varepsilon}
\newcommand{\PKE}{\Pi}
\newcommand{\SKE}{\Sigma}
\newcommand{\GammaSK}{\Gamma^\SKE}
\newcommand{\GammaPK}{\Gamma^\PKE}

\newcommand{\PPT}{{{\rm\sc ppt}}}

\newcommand{\Cont}{\mathrm{Cont}}
\newcommand{\Share}{{\sf Share}}
\newcommand{\Recon}{{\sf Reconstruct}}

\definecolor{commentColor}{rgb}{0,0.6,0.8}
\definecolor{PcommentColor}{rgb}{1,0.64,0}
\newif\ifcomments

\newcommand{\Scomment}[1]{
\ifcomments
\textcolor{commentColor}{\small /*S: #1 */}
\else
{}
\fi}
\newcommand{\Pcomment}[1]{
\ifcomments
\textcolor{PcommentColor}{\small /*P: #1 */}
\else
{}
\fi}

\usepackage{framed}

\fancypagestyle{specialfooter}{%
  \fancyhf{}

  \fancyfoot[L]{\it\small This is a working paper, of which an extended abstract appeared in the 15th ACM Conference on Economics and Computation (EC 2014).}
}

\begin{document}
\commentsfalse

\title{Cryptographically Blinded Games:\\Leveraging Players' Limitations for Equilibria and Profit}
\author[1]{Pavel Hub\'{a}\v{c}ek}
\affil[1]{Aarhus University}
\author[2]{Sunoo Park}
\affil[2]{MIT}
\date{}
\maketitle

\thispagestyle{specialfooter}

\begin{abstract}
In this work we apply methods from cryptography to enable any number of mutually distrusting players
to implement broad classes of mediated equilibria of strategic games without the need for trusted mediation.

Our implementation makes use of a (standard) pre-play ``cheap talk'' phase,
in which players engage in free and non-binding communication prior to playing in the original game.
In our cheap talk phase, the players execute a secure multi-party computation protocol
to sample an action profile from an equilibrium of a ``cryptographically blinded'' version of the original game,
in which actions are encrypted. The essence of our approach is to exploit
the power of encryption
to selectively restrict the information available to players
about sampled action profiles, such
that these desirable equilibria can be stably achieved.
In contrast to previous applications of cryptography to game theory, this work is the first to employ the paradigm of
using encryption to allow players to benefit from hiding information
\emph{from themselves}, rather than from others; and we stress that rational players would \emph{choose} to hide
the information from themselves.
\end{abstract}


\section{Introduction}\label{sec:intro}

Nash equilibrium \cite{nash1950} and correlated equilibrium \cite{aumann1974} are important solution concepts that have been
extensively studied in both traditional and computational game-theoretic contexts. Coarse correlated equilibrium
\cite{moulin1978} is a closely related concept that was proposed as a generalization of correlated equilibrium,
which can be more powerful in some settings such as potential games.

In this work we construct protocols for mutually distrusting players to implement any coarse correlated equilibrium (and therefore any correlated equilibrium)
of a strategic game without trusted mediation, via cryptographic cheap talk protocols.
Our approach draws upon cryptography in two ways: first, we introduce
an intermediate, ``cryptographically blinded'' game from which the players sample according to the desired equilibrium; and second,
this sampling is achieved using a secure multi-party computation protocol.
Our results address both the computational and perfect (information-theoretic) settings.



\paragraph{Correlated equilibrium}
Suppose a mediator samples an action profile $a$ from a known distribution $\alpha$, and gives as ``advice'' to
each player $i$ his action $a_i$ in $a$. The distribution $\alpha$ is a correlated equilibrium if, \emph{having seen his advice},
and believing that all other players will follow their advice, no player has incentive to unilaterally deviate from the advice profile.
\cite{aumann1974} showed that correlated equilibria can achieve higher expected payoffs than Nash equilibria.

\paragraph{Coarse correlated equilibrium}
Coarse correlated equilibria are a generalization of correlated equilibria which invokes a notion of commitment.
In the mediated scenario described above, $\alpha$ is a
coarse correlated equilibrium if no player has incentive not to ``promise'' or ``commit'' in advance
-- \emph{before seeing his advice} $a_i$ -- to play according to the advice,
as long as he believes that all other players will commit to do the same.
Note that if a player does not commit, then he will not see the advice at all, and must therefore play an independent strategy:
this is in contrast to correlated equilibria, where deviations may depend on the received advice.

\cite{moulin2013} showed that there is a class of potential games in which the Nash equilibrium payoffs
can be improved upon by coarse correlated equilibria but not by correlated equilibria
(e.g. the Cournot duopoly and public good provision games).

\begin{example}
Let us give a brief example to illustrate the gap between the two types of equilibria.
Suppose Alice plays a game $\Gamma$ where she has a ``safe strategy'' for which her payoff is always zero.
Let $\alpha$ be a distribution over action profiles of $\Gamma$,
and suppose Alice's expected payoff from $\alpha$ is very high, say, a million dollars --
however, some action profiles from $\alpha$ will give her negative payoff.
Now, when Alice receives her advice from the mediator, she might be able to deduce that her payoff
in the advised action profile will be negative. If this is the case, she will choose to deviate to her safe strategy,
so $\alpha$ is not a correlated equilibrium. However, $\alpha$ may still be a coarse correlated equilibrium if Alice can commit
before seeing her advice; and importantly, $\alpha$ may be very desirable from Alice's (risk-neutral) point of view,
since expected payoff is high.
\end{example}

\subsection{Our results}

In this work we address the following question:

\begin{center}
\emph{How can the players of a strategic game implement any coarse correlated equilibrium via (cryptographic) pre-play communication
without trusting each other or a mediator?}
\end{center}

In the computational setting, we give an implementation for general strategic games, in the form of an extended game comprising
a \emph{cryptographic protocol} in the pre-play phase, which securely samples an action profile for a ``cryptographically blinded''
version of the original game, followed by play in the original game. The blinded game's action space consists of \emph{encryptions}
of the original game's actions.

Our implementation has the strong property that any computational coarse correlated equilibrium of the original game
corresponds to a payoff-equivalent Nash equilibrium of the extended game. Furthermore, it
achieves \emph{strategic equivalence} to the original game,
in that every computational Nash equilibrium of the extended game corresponds to a computational coarse correlated equilibrium of the original game.
Pre-play communication is via broadcast, as is standard in the cheap talk literature.

In the information-theoretic setting, we give an implementation for strategic games with four or more players, using a similar
format of a cryptographically blinded pre-play phase followed by (simultaneous) play in the original game, given private pairwise communication
channels between players.
As in the computational setting, we achieve strategic equivalence.
Both the restriction to four or more players and the need for a stronger communication model than broadcast
are unavoidable, as shown by impossibility results of \cite{barany1992,aumann2003long}
which will be discussed in more detail in the next section.

None of our constructions require trusted mediation. After the pre-play phase is complete, there is a single step in which the players
invoke a \emph{verifiable proxy} to play the original game according to their instructions.
Verifiable parties were introduced in \cite{ILM11}, and will be
detailed further in Section~\ref{subsec:priorWork}.
No trust need be placed in the verifiable proxy,
because anyone can check whether ir has acted correctly; and we stress that
unlike the usual mediator for coarse correlated equilibria,
the verifiable proxy does not communicate anything to the players which may \emph{affect their strategies} in the game.
Informally, it simply performs a ``translation'' of a player's chosen strategy from one form into another.

Finally, our constructions require \emph{no physical assumptions} and can be executed entirely over a distributed network.
This contrasts with a number of previous works such as \cite{LMP04,ILM11} which require ``physical envelopes''.

\subsection{Relation to prior work}\label{subsec:priorWork}

\paragraph{Cheap talk}
The \emph{pre-play} literature considers the general problem of implementing equilibria without mediation,
as follows: given an abstract game $\Gamma$, the aim is to devise
a concrete communication game $\Gamma'$ having an equilibrium that is payoff-equivalent to a desirable equilibrium
in $\Gamma$, where the concrete game may have a pre-play \emph{cheap talk} phase in which players engage in
communication that is neither costly nor binding, and has no impact on players'
payoffs except insofar as it may influence future actions.
In the literature there has been much focus on implementing correlated equilibria
\cite{barany1992,BenPorath1998,aumann2003long}.

\paragraph{Power of commitment} It has long been recognized that the possibility to \emph{commit} to
strategies in advance can increase the payoffs achievable in a game, starting with the work of \cite{von1934marktform},
who proposed a leader/follower structure to games where the leader moves first (and thereby ``commits'' to his strategy).
\cite{DBLP:journals/geb/StengelZ10} showed that transforming a strategic game into a leader/follower form
allows the leader (i.e. the committer) to do at least as well as in the Nash and correlated equilibria of the strategic game.
Moreover, they show that coarse correlated equilibria, with their arguably stronger notion of commitment,
can yield higher payoffs than the leader/follower transformation.
More recently, \cite{letchford2012value} studied the advantage of commitment from a quantitative perspective
and showed that the extremal ``value of commitment'' is in fact unbounded in many classes of games.

In this work, we achieve the payoffs of coarse correlated equilibria without resorting to the assumption of
binding contracts: instead, we use the power of encryption to hide information that, if known to the
players, could render the situation unstable. We stress that the players are \emph{given the choice}, rather than forced,
to hide information from themselves -- and we find that it is in their rational interest to do so since coarse correlated
equilibria can offer high payoffs.

\paragraph{Cryptographic cheap talk and computational equilibria}
\cite{DHR00} introduced the idea of \emph{cryptographic cheap talk}, in which
players execute a cryptographic protocol during the pre-play phase; and they defined \emph{computational equilibria},
which are solution concepts stable for computationally bounded (probabilistic polynomial time) players who are
indifferent to negligible gains. Their cryptographic cheap talk protocols efficiently implement some computational correlated equilibria
of two-player games. Moreover, their notion of computational equilibria
suffers from \emph{empty threats}
(Definition~\ref{def:emptyThreatInformal}), which cause instability for sequentially rational players in the pre-play game.
This was partially addressed by a new solution concept of \cite{GLR10}; however,
\cite{HNR13} subsequently showed that in general, correlated equilibria cannot be achieved without empty threats by (cryptographic) cheap talk.

Our results in the computational setting use the equilibrium definitions of \cite{DHR00};
however, in our ``cryptographically blinded'' games, empty threats cannot occur.
By converting games into blinded games, our constructions implement all
coarse correlated equilibria without empty threats: this comes at the cost of a single mediated ``translation'' step using a
third party,
discussed in the next paragraph. We consider this step to be a ``necessary'' and mild requirement given that
the impossibility result of \cite{HNR13} renders some additional assumption necessary to achieve all (coarse) correlated equilibria
without empty threats.

\paragraph{Removing trusted mediation}
Removing the need to trust a mediator in the implementation of equilibria and mechanisms has long been a subject of
interest in game theory and cryptography. The notion of \emph{verifiable mediation} was introduced by \cite{ILM11},
who highlighted the difference between the usual concept of a \emph{trusted mediator}, and the weaker
concept of a \emph{verifiable mediator} who performs actions in a publicly verifiable way and without possessing any information that should be kept secret.
Recent applications of verifiable mediation include the strong correlated equilibrium implementation of \cite{ILM11},
and the rational secret sharing scheme of \cite{MS09}.

In this paper, we introduce the new notion of a \emph{verifiable proxy}.
As in verifiable mediation, the actions of a verifiable proxy are publicly verifiable. However, our notion is incomparable to \cite{ILM11}'s
verifiable mediation, because:
\begin{itemize}
\item a verifiable proxy for a strategic game does not give the players any information that affects their strategic choices in the game; and
\item a verifiable proxy may possess information that should be kept secret.
\end{itemize}
More discussion about the merits of these definitions is given in Section~\ref{subsec:whatCanIDo}.

As a simple illustration, consider a sealed-bid auction: much more trust is placed in a mediator who collects
all the players' bids and just announces the winner, than in a mediator who collects the bids, opens them publicly,
and allows everyone to compute the outcome themselves.

In our setting, the verifiable proxy performs a single ``translation'' step on behalf of the players, at the end of the pre-play phase,
in which it takes strategies submitted by the players and ``translates'' them into a different format.
In particular, the proxy acts independently and identically with respect to each player, and therefore is
not implementing the correlation aspect.

\paragraph{Strategic equivalence property}
An important concern in implementation theory is the strategic equivalence of an implementation to the underlying game:
it is desirable that implementations have the ``same'' equilibria as the underlying game, and in particular do not
introduce new ones.
This was first considered by the \emph{full implementation} concept of \cite{maskin1999}, and extended by subsequent works
such as \cite{ILM11} who proposed a stronger notion of \emph{perfect implementation} for certain games.
Although this literature is not directly applicable to the present work (as our results lie in the pre-play realm),
we extend these ideas and find that the cheap talk extensions of our ``cryptographically blinded'' games
achieve full strategic equivalence in that their Nash equilibria correspond exactly to the coarse correlated equilibria of their underlying games.

\paragraph{Computationally unbounded setting} To our knowledge, existing work in applying cryptographic tools to game theory has
focused overwhelmingly on the setting of computationally bounded players and computational equilibria. In contrast, we consider
the computationally unbounded setting too. Our result for the computational setting is stronger and more efficient than our
information-theoretic solution: in particular, the computational result holds for games with any number of players,
and requires only a broadcast channel for communication between players.

In the computationally unbounded setting it was proven by \cite{barany1992} that correlated equilibria cannot be
achieved by cheap talk between fewer than four players, and indeed, this fits neatly with a more general result of \cite{BGW88,CCD88}
in the context of secure protocols.
Accordingly, our information-theoretic results only apply for games of four or more players; however, improving on the protocols of
\cite{barany1992}, we achieve not only correlated equilibria but coarse correlated equilibria for all games of this type.

Furthermore, in the computationally unbounded setting it has been proven \cite{aumann2003long} that communication by broadcast alone is \emph{insufficient} to achieve
(non-trivial) correlated equilibria by cheap talk, so our result is of interest notwithstanding its stronger requirement of
private communication channels between players. Indeed, the private-channels model has been extensively studied
in both distributed computing (e.g. \cite{FLP85,gossip03}) and multi-party computation (e.g. \cite{BGW88,CCD88}) as an interesting strengthening of the communication model
that allows for much stronger and/or more efficient protocols than the broadcast model.
We therefore consider it natural and compelling to apply this model in the game-theoretical setting.

\subsection{Organization}
In Sections~\ref{sec:gameTheory} and \ref{sec:crypto} we provide game-theoretical and cryptographic background.
In Section~\ref{sec:blindedGames} we introduce cryptographically blinded games. These are
the essential building block for the cheap talk protocols detailed in Section~\ref{sec:protocols} that
implement all coarse correlated equilibria of general strategic games. At the end of Section~\ref{sec:protocols}
we discuss the efficiency of our protocols.

\subsection{Notation}
For $n\in\NN$, let $[n]$ denote the set $\{1,2,\dots,n\}$.
For a set $S$, let $\powset(S)$ denote the powerset of $S$, and let $\Delta(S)$ denote the set of all distributions over $S$.
Let $s\larr S$ denote that $s$ is a random element of $S$.
Let $\sqcup$ denote the disjoint union operation.
We write \PPT{} to mean probabilistic polynomial time, and we call distributions that can be sampled in probabilistic polynomial time
``\PPT-samplable''.
Let $\negl$ denote a negligible function (which tends to zero faster than any inverse polynomial).

\section{Game-theoretic background}\label{sec:gameTheory}

\begin{definition}[Finite strategic game]
A finite strategic game $\Gamma=\langle N,(A_i),(u_i)\rangle$ is defined by
a finite set $N$ of players, and
for each player $i\in N$, a non-empty set of possible actions $A_i$ and
a utility function $u_i:\times_{j\in N}A_j\rarr\RR$.
\end{definition}

We refer to an \emph{action profile}
$a=(a_j)_{j\in N}$ of a game as an \emph{outcome}, and denote by $A$ the set of outcomes $\times_{j\in N}A_j$.
For a given outcome $a$, we write $a_{-i}$ to denote $(a_j)_{j\in N, j\neq i}$, that is, the
profile of actions of all players other than $i$; and we use $(a'_i,a_{-i})$ to denote the action
profile where player $i$'s action is $a'_i$ and all other players' actions are as in $a$.

\subsection{Equilibrium concepts}

\begin{definition}[Nash equilibrium]\label{def:NE}
A \emph{Nash equilibrium} of strategic game $\Gamma=\langle N,(A_i),(u_i)\rangle$
is a product distribution $\alpha\in\times_{j\in N}\Delta(A_j)$ such that
for every player $i\in N$ and for all $a^*_i\in A_i$
\[
	\E_{a\larr\alpha}[u_i(a)]
		\geq \E_{a\larr\alpha}[u_i(a^*_i,a_{-i})].
\]
\end{definition}

\begin{definition}[Correlated equilibrium]\label{def:CE}
A \emph{correlated equilibrium} of strategic game $\Gamma=\langle N,(A_i),(u_i)\rangle$
is a probability distribution $\alpha\in\Delta(\times_{j\in N}A_j)$ such that
for every player $i\in N$, and for all $b_i,a^*_i\in A_i$ satisfying $\Pr_{a\larr\alpha}[a_i=b_i]>0$,
\[
	\E_{a\larr\alpha}[u_i(a)|a_i=b_i]
		\geq \E_{a\larr\alpha}[u_i(a^*_i,a_{-i})|a_i=b_i].
\]
\end{definition}

\begin{definition}[Coarse correlated equilibrium]\label{def:CCE}
A \emph{coarse correlated equilibrium} of strategic game $\Gamma=\langle N,(A_i),(u_i)\rangle$
is a probability distribution $\alpha\in\Delta(\times_{j\in N}A_j)$ such that
for every player $i\in N$ and for all $a^*_i\in A_i$
\[
	\E_{a\larr\alpha}[u_i(a)]
		\geq \E_{a\larr\alpha}[u_i(a^*_i,a_{-i})].
\]
\end{definition}

The model of coarse correlated equilibrium allows the players either to ``commit in advance''
to play according to the mediator's advice (no matter what it turns out to be),
or to play an \emph{independent} strategy without learning the advice.
A probability distribution is a coarse correlated equilibrium if no player has
an incentive to not commit to play according to the mediator's advice.

Because of linearity of expectation, it is sufficient for these
equilibrium definitions to consider only deviations to pure strategies.
Note that any Nash equilibrium is a correlated equilibrium, and any
correlated equilibrium is a coarse correlated equilibrium.

\subsection{Computational equilibrium concepts}\label{sec:computationalEquilibria}

The following definitions of computational equilibria extend those introduced by \cite{DHR00}.
In the computational setting a strategic game induces a family of games parametrized by the security parameter,
i.e.  $\Gamma=\{\langle N,(A_{i}^{(k)}),(u_{i}^{(k)})\rangle\}_{k\in\NN}$.
Hence, the corresponding solution concepts are ensembles of probability distributions,
and the security parameter captures the intuition that players are limited to efficiently computable
(\PPT{}) strategies and indifferent to gains negligible in $k$.


\begin{definition}[Computational Nash equilibrium] \label{def:compNE}
A \emph{computational Nash equilibrium} of computational strategic game
$\Gamma=\{\langle N,(A_{i}^{(k)}),(u_{i}^{(k)})\rangle\}_{k\in\NN}$
is a \PPT-samplable ensemble of product distributions $\alpha=\{\alpha^{(k)}=\times_{j\in N}\alpha_j^{(k)}\}_{k\in\NN}$
on $\{\times_{j\in N}A_j^{(k)}\}_{k\in\NN}$ such that for all players $i\in N$ and every \PPT-samplable ensemble $\hat{\alpha}_i=\{\hat{\alpha}_{i}^{(k)}\}_{k\in\NN}$
on $\{A_i^{(k)}\}_{k\in\NN}$, there exists a negligible $\eps(\cdot)$ such that for all large enough $k\in\NN$ it holds that
\[
	\E_{a\larr\alpha^{(k)}}[u_{i}^{(k)}(a)]
		\geq \E_{a\larr\alpha^{(k)},\hat{a}_i\larr\hat{\alpha}_i^{(k)}}[u_{i}^{(k)}(\hat{a}_i,a_{-i})]-\eps(k).
\]
\end{definition}

\begin{definition}[Computational correlated equilibrium]\label{def:compCE}
A \emph{computational correlated equilibrium} of computational strategic game $\Gamma=\{\langle N,(A_{i}^{(k)}),(u_{i}^{(k)})\rangle\}_{k\in\NN}$
is a \PPT-samplable probability ensemble $\alpha=\{\alpha^{(k)}\}_{k\in\NN}$ on $\{\times_{j\in N}A_j^{(k)}\}_{k\in\NN}$
such that for all players $i\in N$ and every \PPT-samplable ensemble $\hat{\alpha}_i=\{\hat{\alpha}^{(k)}_{i}\}_{k\in\NN}$
on $\{A_i^{(k)}\}_{k\in\NN}$ there exists a negligible $\eps(\cdot)$ such that for all large enough $k\in\NN$ it holds that
\[
	\E_{a\larr\alpha^{(k)}}[u_{i}^{(k)}(a)]
		\geq \E_{a\larr\alpha^{(k)},\hat{a}_i\larr\hat{\alpha}_{i}^{(k)}(a_i)}[u_{i}^{(k)}(\hat{a}_i,a_{-i})]-\eps(k).
\]
\end{definition}

\begin{definition}[Computational coarse correlated equilibrium] \label{def:compCCE}
A \emph{computational coarse correlated equilibrium} of computational strategic game
$\Gamma=\{\langle N,(A_{i}^{(k)}),(u_{i}^{(k)})\rangle\}_{k\in\NN}$ is a \PPT-samplable probability ensemble $\alpha=\{\alpha^{(k)}\}_{k\in\NN}$
on $\{\times_{j\in N}A_j^{(k)}\}_{k\in\NN}$ such that for all players $i\in N$ and every \PPT-samplable ensemble $\hat{\alpha}_i=\{\hat{\alpha}_{i}^{(k)}\}_{k\in\NN}$
on $\{A_i^{(k)}\}_{k\in\NN}$, there exists a negligible $\eps(\cdot)$ such that for all large enough $k\in\NN$ it holds that
\[
	\E_{a\larr\alpha^{(k)}}[u_{i}^{(k)}(a)]
		\geq \E_{a\larr\alpha^{(k)},\hat{a}_i\larr\hat{\alpha}_i^{(k)}}[u_{i}^{(k)}(\hat{a}_i,a_{-i})]-\eps(k).
\]
\end{definition}

Note that in the above definition of computational coarse correlated equilibrium the output of $\hat{\alpha}_{i}^{(k)}$
is independent of $a_i$, unlike in the definition of computational correlated equilibrium.

%

\begin{remark}
In later sections we apply the above computational solution concepts in a straightforward way to classical strategic games.
For a finite strategic game $\Gamma=\langle N, (A_i),(u_i)\rangle$ we consider the computational version $\{\Gamma^{(k)}\}_{k\in\NN}$,
where $\Gamma^{(k)}=\Gamma$ for all $k\in\NN$.
The action space and the utility function do not change with the security parameter in this computational version of $\Gamma$;
however, the players are limited to efficient (\PPT{}) strategies.

\end{remark}

\begin{remark}
In the classical setting, it is implicit that the players of a game have oracle access to the utility functions $u_i$,
that is, players can query $u_i$ on any action profile in constant time\footnote{Other parameters of the original game,
such as the correlated equilibrium distribution, are also assumed to be computable in constant time.}.
Our results apply to all strategic games in the classical setting: hence the requirement
that the security parameter be polynomial in the size of the game (i.e. we ensure that players
are able to perform the standard task of reading the payoff matrix).
With computationally bounded players, however, it seems very natural to consider
the case in which computing $u_i$ takes more time.
To our knowledge, this difference has been recognized (e.g., \cite{DHR00}) but
not much analyzed in the literature; however, it is an important underlying idea
of the present work.
\end{remark}

\subsection{Extensive games}

Definitions of extensive form games and subgames are given in Appendix~\ref{appx:extensiveGames}, along with
corresponding equilibrium concepts for the standard and computational settings.

\section{Cryptographic background}\label{sec:crypto}

\subsection{Encryption schemes}

Our constructions will make use of secret-key and public-key encryption schemes, which are defined below.
Note that encryption schemes are parametrized by a security parameter $k$
that determines the ``security level'' of the scheme.

\begin{definition}[Secret-key encryption scheme]
A \emph{secret-key encryption scheme} over a message space $\Msg$ is a tuple of
\PPT{} algorithms $\SKE=(\SGen,\SEnc,\SDec)$ satisfying the following.
Let the ciphertext space be the codomain of $\SEnc$ and be denoted by $\Ciph$.
\begin{itemize}
\item The key generation algorithm $\SGen$ takes no input and outputs a secret key $sk$ according to
some distribution (inherent to $\SKE$). This is denoted by $sk\larr\SGen()$.
\item The encryption algorithm $\SEnc$ takes as input a message $m\in\Msg$ and a secret key $sk$, and outputs
a ciphertext $c\in\Ciph$. This is denoted by $c\larr\SEnc_{sk}(m)$.
\item The decryption algorithm $\SDec$ is a deterministic algorithm that takes as input a ciphertext $c$
and a secret key $sk$, and outputs a decryption $m'\in\Msg$. This is denoted by $m'=\SDec_{sk}(c)$.
\item The decryption is always correct, i.e.
		for every security parameter $k$, and every $sk\larr\SGen()$
		it holds for every $m\in\Msg$ that $\SDec_{sk}(\SEnc_{sk}(m))=m$.
\end{itemize}
\end{definition}

\begin{definition}[Public-key encryption scheme]
A \emph{public-key encryption scheme} over a message space $\Msg$ is a tuple of \PPT{} algorithms
$\PKE=(\PGen,\PEnc,\PDec)$ satisfying the following.
Let the ciphertext space be the codomain of $\PEnc$ and be denoted by $\Ciph$.
\begin{itemize}
	\item The key generation algorithm $\PGen$ takes input $1^k$, where $k$ is the security parameter, and outputs
		a public key and secret key pair $(pk,sk)$. 
	\item The encryption algorithm $\PEnc$ takes as input a message $m\in\Msg$
		and a public key $pk$ and outputs a ciphertext $c\in\Ciph$. 
	\item The decryption algorithm $\PDec$ is a deterministic algorithm that takes
		as input a ciphertext $c$ and a secret key $sk$, and outputs a decryption $m'\in\Msg$.
	\item The decryption is always correct, i.e.
		for every security parameter $k$, and every $(pk,sk)\larr\PGen(1^k)$
		it holds for every $m\in\Msg$ that $\PDec_{sk}(\PEnc_{pk}(m))=m$.
\end{itemize}
\end{definition}


\subsection{Security definitions}

Here we define the following two standard security notions: perfect (information-theoretic) security,
and computational security against chosen-ciphertext attacks. The latter is commonly referred to as $\CCA$-security,
and is the de facto standard for security of public-key encryption; the former is canonical in the information-theoretic setting.

\begin{remark}
Our constructions make use of perfectly secure secret-key encryption and $\CCA$-secure public-key encryption. For convenience,
therefore, the security definitions given below refer to secret- and public-key schemes respectively.
However, both security definitions may be straightforwardly adapted to apply to both types of encryption
(although it is well known that perfect security is impossible in the public-key setting).
\end{remark}

\begin{definition}[Perfectly secure secret-key encryption]\label{def:perfectSecurity}
A secret-key encryption scheme $\SKE=(\SGen,\SEnc,\SDec)$ is \emph{perfectly secure} if
for all messages $m_0,m_1\in\Msg$
and ciphertexts $c\in\Ciph$, it holds that
$\Pr[\SDec(\SEnc(m_0))=m_0]=1$
and
$$\Pr_{sk\larr\SGen()}[\SDec_{sk}(c)=m_0] = \Pr_{sk\larr\SGen()}[\SDec_{sk}(c)=m_1].$$
\end{definition}

An alternative and equivalent definition is that a perfectly secure encryption scheme produces ciphertexts that are independent of the messages that they encrypt.

Next, we shall define $\CCA$-security for public-key encryption schemes. The security definition is based on the following experiment,
which may be considered to be a game played between a malicious adversary $\Adv$ and an honest challenger.

\begin{framed}
\begin{center}
The $\CCA$ indistinguishability experiment $\PubK_{\Adv,\PKE}^{\CCA}(k)$:
{\small \begin{enumerate}
	\item The challenger generates a key pair $(pk,sk)\larr\PGen(1^k)$, and sends $(1^k,pk)$ to $\Adv$.
	\item $\Adv$ has oracle access to $\PDec_{sk}$, and outputs messages $m_0,m_1\in\Msg$ of the same length.
	\item The challenger samples $b\dollarr \{0,1\}$, then computes $c\larr\PEnc_{pk}(m_b)$, and sends $c$ to $\Adv$.
	\item $\Adv$ still has oracle access to $\PDec_{sk}$, but cannot query $\PDec_{sk}(c)$. $\Adv$ now outputs a bit $b'$.
	\item The output of the experiment is $1$ if $b' = b$, and $0$ otherwise.
\end{enumerate} }
\end{center}
\end{framed}

Informally, the adversary ``wins the game'' if he guesses correctly
which of the two messages was encrypted. Clearly, he can win with probability $1/2$
by random guessing. The definition of $\CCA$-security formalizes the intuition that he should not be able to do better than that.

\begin{definition}[$\CCA$-secure public-key encryption]\label{def:CCAsec}
A public-key encryption scheme $\PKE=(\PGen,\PEnc,\PDec)$ is \emph{$\CCA$-secure}
(i.e. secure against chosen-ciphertext attacks), if for all \PPT{} adversaries $\Adv$,
$\Pr[\PubK_{\Adv,\PKE}^{\CCA}(k) = 1] \leq 1/2 + \eps(k)$ for some negligible $\eps$.
\end{definition}

%
%

\subsection{Non-malleable encryption}

Non-malleable encryption was introduced by \cite{DDN00} in the
computational setting, and extended to the information-theoretic setting by \cite{HSHI02}.
Informally, non-malleability requires that given a ciphertext $c$, an adversary (who does not know the secret key
or the message encrypted by $c$) cannot generate a different ciphertext $c'$ such that the
respective messages are related by some known relation $R$.

We begin with the simpler information-theoretic definition.
Note that \cite{HSHI02} also give a construction of perfectly non-malleable secret-key encryption.

\begin{definition}[Perfect non-malleability]\label{def:perfNML}
A secret-key encryption scheme $\SKE=(\SGen,\SEnc,\SDec)$ is \emph{perfectly non-malleable}
if for all $c,c',c''\in\Ciph$ such that $c'\neq c\neq c''$ and all relations $R:\Msg\times\Msg\rarr\{0,1\}$,
$$\Pr_{sk\larr\SGen()}[R(\SDec(c),\SDec(c'))=1] = \Pr_{sk\larr\SGen()}[R(\SDec(c),\SDec(c''))=1].$$
\end{definition}


Observe that perfect non-malleability implies perfect security (but not vice versa).

The computational definition of non-malleability is more involved, using an indistinguishability experiment similar to that of the
$\CCA$-security definition.
It formalizes the same idea, that an attacker
must be unable (with more than negligible advantage) to modify ciphertexts such that the new decryption satisfies a known relation
with the original decryption.
The definition of non-malleability for (public-key) encryption schemes is based on the following experiment.

\begin{framed}
\begin{center}
The $\NM$ indistinguishability experiment $\PubK_{\Adv,\PKE}^{\NM}(k)$:
{\small \begin{enumerate}
	\item The challenger generates a key pair $(pk,sk)\larr\PGen(1^k)$ and sends $(1^k,pk)$ to $\Adv$.
	\item $\Adv$ has oracle access to $\PDec_{sk}$, and outputs (a description of) an efficiently samplable distribution $M$
on the message space $\Msg$ (which must give non-zero probability only to strings of a given length).
	\item The challenger samples a message $m\larr M$, and sends ciphertext $c=\PEnc_{pk}(m)$ to $\Adv$.
	\item $\Adv$ still has oracle access to $\PDec_{sk}$, but cannot query $\PDec_{sk}(c)$. $\Adv$ outputs a ciphertext $c'$ and (a description of) an efficiently computable relation $R:\Msg\times\Msg\rarr\{0,1\}$.
	\item The output of the experiment is $1$ if $c'\neq c$ and $R(m,\PDec_{sk}(c'))$ is true, and $0$ otherwise.
\end{enumerate} }
\end{center}
\end{framed}

Define $\PubK_{\Adv,\PKE}^{\NM,\$}(k)$ to be identical to $\PubK_{\Adv,\PKE}^{\NM}(k)$, except that item 3 is replaced by:
\begin{framed}
{\small \begin{itemize}
\item[$3'$.] The challenger samples independent messages $m,\tilde{m}\larr M$, and sends $c=\PEnc_{pk}(\tilde{m})$ to $\Adv$.
\end{itemize}}
\end{framed}

\begin{definition}[Computationally non-malleable encryption] \label{def:compNM}
A public-key encryption scheme $\PKE=(\PGen,\PEnc,\PDec)$ is \emph{$\NM$-$\CCA$-secure} (that is, non-malleable
against chosen ciphertext attacks), if for all \PPT{} adversaries $\Adv$ there exists a
negligible function $\negl$ such that
\[
\left| \Pr[\PubK_{\Adv,\PKE}^{\NM}(k) = 1] - \Pr[\PubK_{\Adv,\PKE}^{\NM,\$}(k)=1] \right| \leq \negl(k).
\]
\end{definition}

In our setting\footnote{When considering security notions other than $\CCA$, standard indistinguishability-based security does not imply non-malleability.
In this work we only use $\CCA$-secure schemes.}, $\CCA$-security is equivalent to computational non-malleability, as stated in Claim~\ref{claim:nm}.
For the proof, we refer the reader to \cite{BDPR98}.

\begin{claim}\label{claim:nm}
An encryption scheme is $\CCA$-secure (Definition~\ref{def:CCAsec}) if and only if it satisfies
computational non-malleability (Definition~\ref{def:compNM}).
\end{claim}

\subsection{Secure multi-party computation}\label{subsec:mpc}

Consider $N$ players, each with an input value $x_i$ for $i\in N$, who wish to
jointly compute a function $f$ on their inputs: $f(x_1,\cdots,x_N)=(y_1,\dots,y_N)$. 
These players do not trust each other: they want each player $i$ to receive his output value $y_i$ at the end of the computation,
but they also want a guarantee that no player $i$ can learn any information beyond his designated output $y_i$ (even if he ``cheats'').
Multi-party computation gives interactive $N$-party protocols to solve this problem,
with security and correctness guarantees even when some players may maliciously deviate from the protocol.

\begin{definition}[Secure multi-party computation]\label{def:MPC}
An $N$-party computation protocol is said to be \emph{perfectly secure}
(for up to $t<N$ corruptions) if it satisfies the following properties, against any adversary who corrupts
up to $t$ players\footnote{The corrupted players may be thought of as ``dishonest'' players trying to sabotage the protocol.}:
\begin{itemize}
\item \emph{Correctness}: The output of the computation is equal to $f(x_1,\cdots,x_N)$.
\item \emph{Privacy}: No adversary can obtain any information about the honest parties' inputs, 
other than what can be deduced from the corrupted players' input and output values $\{x_i,y_i\}_{i\in S}$
(where $S$ denotes the set of corrupt players).
\end{itemize}

The protocol is said to be \emph{computationally secure} if it satisfies the above properties with all but negligible
probability (in a security parameter $k$) against \PPT{} adversaries.
\end{definition}

The following are general possibility results for multi-party computation that are relevant to this work.
For proofs, we refer the reader to the original papers.\footnote{Dodis and Rabin~\cite{dodis2007cryptography} provide and extended summary of the multi-party computation results with emphasis on the use in the game theoretical context.}

\begin{theorem}[\cite{BGW88,CCD88}]\label{thm:perfectMPC}
Any circuit can be evaluated by an $N$-party protocol with perfect security against $t<N/3$ corruptions.
Moreover, the bound of $t<N/3$ is tight.
\end{theorem}

\begin{theorem}[\cite{GMW87,DBLP:journals/iacr/AsharovL11}]\label{thm:compMPC}
Any circuit can be evaluated by an $N$-party protocol with computational security against up to $t=N-1$ corruptions.
\end{theorem}

An additional desirable property of multi-party computation protocols, other than correctness and privacy, 
is \emph{guaranteed output delivery}: the property that every honest (non-corrupt) player is guaranteed to receive her correct output,
even in the presence of an adversary. This property is known to be achievable if and only if $t<N/2$ 
(that is, a majority of the players are honest) \cite{GMW87,Cle86}.

\subsection{Secret Sharing}

A secret sharing scheme specifies a method for a special party (the ``dealer'')
to \emph{share} a secret $s$ among $N$ players so that only large enough subsets of players can reconstruct the secret value $s$.
The dealer gives privately a share $s_i$ to each player $i$, so that
any set of up to $k-1$ shares contains no information about $s$;
however, it can efficiently be reconstructed given any $k$ or more shares.
The formal definition is given below.

\begin{definition}[Secret sharing scheme \cite{Shamir:1979:SS:359168.359176}]
A \emph{$k$-out-of-$N$ secret sharing scheme} is a pair of algorithms $(\Share,\Recon)$ as follows.
$\Share$ takes as input a secret value $s$ and outputs a set of \emph{shares} $S=\{s_1,\dots,s_N\}$ such that
the following two properties hold.
\begin{itemize}
\item \emph{Correctness}: For any subset $S'\subseteq S$ of size $|S'|\geq k$, it holds that $\Recon(S')=s$, and
\item \emph{Privacy}: For any subset $S'\subseteq S$ of size $|S'|< k$, it holds that $H(s)=H(s|S')$, where $H$ denotes the binary entropy function.
\end{itemize}
$\Recon$ takes as input a (sub)set $S'$ of shares and outputs:
$$\Recon(S')=\begin{cases}\bot & \mbox{if} \qquad |S'|<k \\ s & \mbox{if} \qquad\exists S \mbox{ s.t. } S'\subseteq S \mbox{ and } \Share(s)=S \mbox{ and } |S'|\geq k \end{cases}.$$
\end{definition}

\section{Cryptographically blinded games}\label{sec:blindedGames}

Now we define ``cryptographically blinded'' games $\Gamma'$ whose
actions are encryptions of the actions of an underlying game $\Gamma$.
Payoffs from corresponding action profiles of $\Gamma$ and $\Gamma'$ are the same.
These blinded games will be an essential tool for our pre-play protocols, which will be detailed in Section~\ref{sec:protocols}.

The following supporting definition formalizes the intuitive notion that two strategic games
are equivalent up to renaming of actions or deletion of redundant actions.

\begin{definition} \label{def:equiv_game}
For any strategic game $\Gamma=\langle N,(A_i),(u_i)\rangle$,
a strategic game $\Gamma'=\langle N,(A_i'),(u_i')\rangle$ is said to be \emph{super-equivalent} to $\Gamma$
if there exist surjective \emph{renaming functions} $\rho_i:A_i'\rarr A_i$ such that for all $i\in N$, for all $a_1'\in A_1',\dots,a_N'\in A_N'$, it holds that
$u_i'(a_1',\dots,a_N')=u_i(\rho_1(a_1'),\dots,\rho_N(a_N'))$.
In this case, we write $\Gamma'\eqvrho\Gamma$.
\end{definition}

\paragraph{Notation} For a renaming function $\rho$, let $\rhoinv_i:A_i\rarr\powset(A_i')$ be defined by $\rhoinv_i(a_i)=\{a'_i|\rho(a'_i)=a_i\}$. To simplify notation, we define $\rho:A_1\times\dots\times A_N\rarr A_1'\times\dots\times A_N'$ to be $\rho(a_1,\dots,a_N)=(\rho_1(a_1),\dots,\rho_N(a_N))$, and let $\rhoinv$ be defined similarly.
For a distribution $\gamma'$ on action profiles of $\Gamma'$,
$\rho(\gamma')$ denotes the distribution on action profiles of $\Gamma$
that corresponds to sampling $a'\in A'$ according to $\gamma'$
and outputting $\rho(a')$.


\begin{lemma}\label{lem:rhoEquivalenceOfCCE}
	Let $\Gamma$ be a strategic game. Then for any $\Gamma'$ with
	$\Gamma'\eqvrho\Gamma$ it holds that: (1) for any coarse correlated equilibrium $\alpha$ of $\Gamma$,
			there exists a coarse correlated equilibrium $\alpha'$ of $\Gamma'$
			such that $\rho(\alpha')=\alpha$; and (2) for any coarse correlated equilibrium $\alpha'$of $\Gamma'$,
			$\rho(\alpha')$ is a coarse correlated equilibrium of $\Gamma$.
\end{lemma}

\begin{proof}
To show item (1), consider the distribution $\alpha'$
on action profiles of $\Gamma'$ obtained by
sampling an action profile $a$ from $\alpha$ and outputting a random $a'\in\rho^{-1}(a)$.
Note that $\rho(\alpha')=\alpha$ by construction.	
We need to show that for all $i\in N$ and all $a^*_i\in A'_i$,
\[
\E_{a'\larr\alpha'}[u'_i(a')]
\geq \E_{a'\larr\alpha'}[u'_i(a^*_i,a'_{-i})].
\]
The above can be rewritten, due to the construction of $\alpha'$ and definition of $\Gamma'$, as
$
\E_{a\larr\rho(\alpha')}[u_i(a)]
\geq \E_{a\larr\rho(\alpha')}[u_i(\rho_i(a^*_i),a_{-i})]
$.
This holds for every $\rho_i(a^*_i)\in A_i$ since $\alpha=\rho(\alpha')$
is a coarse correlated equilibrium of $\Gamma$.
Item (2) follows similarly, since $\rho(\alpha')$
is a distribution on action profiles of $\Gamma$ and
$\alpha'$ is a coarse correlated equilibrium.
\end{proof}

The interesting case of the seemingly straightforward definition of super-equivalence
arises when the renaming function $\rho$ is not invertible by the players.

We now define cryptographically blinded games.
Let $\Gamma=\langle N,(A_i),(u_i)\rangle$ be a strategic game,
where players have oracle access to the utility functions $u_i$.

\begin{definition}[Secret-key blinded game]\label{def:PKgame}
Let $\SKE=(\SGen,\SEnc,\SDec)$ be a secret-key encryption scheme, and let $\Gamma=\langle N,(A_i),(u_i)\rangle$ be a strategic game.
Define the \emph{blinded game} $\GammaSK=\langle N,(A_i'),(u_i)'\rangle$ of $\Gamma$ to be the game such that $sk\larr\SGen()$ is generated and
\begin{itemize}
	\item for each player $i\in N$ the action space is $A_i'=A_i\sqcup\{\SEnc_{sk}(a_i)|a_i\in A_i\}$
	\item for each player $i\in N$ the utility for all $a'\in\times_{j\in N} A'_j$ is $u_i'(a')=u_i(a)$, where for all $j\in N$
	\[
	a_j=
	\begin{cases}
	a'_j							&\text{if } a'_j\in A_j,\\
	\SDec_{sk}(a'_j)  &\text{otherwise.}
  \end{cases}
	\]
\end{itemize}
 \end{definition}

\begin{definition}[Public-key blinded game]\label{def:PKgame}
Let $\PKE=(\PGen,\PEnc,\PDec)$ be a public-key encryption scheme, and let $\Gamma=\langle N,(A_i),(u_i)\rangle$ be a strategic game.
Define the \emph{computational blinded game} $\GammaPK=\{\langle N,(A_{i}^{\prime{}(k)}),(u_{i}^{\prime{}(k)})\rangle\}_{k\in\NN}$ of $\Gamma$ to be the computational game such that for every security parameter $k\in\NN$ a corresponding key pair $(pk,sk)\larr\PGen(1^k)$ is generated and
\begin{itemize}
	\item for each player $i\in N$ the action space is $A_{i}^{\prime{}(k)}=\{\PEnc_{pk}(a_i)|a_i\in A_i\}$
	\item for each player $i\in N$ the utility for all $a'\in\times_{i\in N} A_{i}^{\prime{}(k)}$ is $u_{i}^{\prime{}(k)}(a')=u_i(\PDec_{sk}(a'))$.
\end{itemize}
\end{definition}

If $\GammaPK$ and $\GammaSK$ are blinded games of the game $\Gamma$, then we say that
$\Gamma$ is the \emph{underlying game} of $\GammaPK$ and $\GammaSK$.

Observe that the blinded games $\GammaPK$ and $\GammaSK$ are super-equivalent to the underlying game $\Gamma$, with respect to
renaming functions $\rho=\PDec_{sk}$ or $\rho=\SDec_{sk}$ (respectively).

\begin{remark}
In these contexts, players do not have knowledge of
the secret key $sk$, as is standard and necessary when employing encryption schemes.
Therefore, expectations ``from the point of view of the player'' are taken over a secret key $sk\larr\SGen()$ or $(pk,sk)\larr\PGen(1^k)$, where
secret- or public-key encryption schemes are used, respectively.
\end{remark}

It is assumed to be infeasible for players of a game $\Gamma$ to efficiently compute
the utility functions $u_i'$ on arbitrary action profiles in $\GammaSK$ or $\GammaPK$, since they cannot (efficiently) decrypt ciphertexts
in the corresponding encryption schemes.
However, our applications require players to be able to pick actions in $A_i'$
for which they know the corresponding expected utility. In fact, if the players cannot do this, then the games become
meaningless in that \emph{any} distribution on $A$ is an equilibrium. In the public-key case, this property is achieved
as players can simply compute the encryption of some $a_i\in A_i$ for which the utility is known.
In the secret-key case, $A_i$ is contained in $A_i'$ for exactly this purpose.

\paragraph{Security parameter for public-key games}
Public-key blinded games have an implicit security parameter $k$ due to
the underlying encryption scheme. When applying computational equilibrium concepts (which have a security parameter $k'$ of their own)
to such games, there must be a fixed relation between $k$ and $k'$ in order to have a meaningful definition of security
for a computational equilibrium of a blinded game.
In our setting, both parameters represent the same quantity:
the computational boundedness of the players of a game. Therefore, we let $k=k'$ and refer to a single security parameter $k$.

\subsection{Correspondence of equilibria in blinded games}

\begin{lemma}\label{lem:CCEvsCEinSKE}
Let $\SKE=(\SGen,\SEnc,\SDec)$ be a perfectly non-malleable and verifiably decryptable secret-key encryption scheme.
Then for any strategic game $\Gamma$,
it holds that for any coarse correlated equilibrium $\alpha$ of $\Gamma$ there exists a
correlated equilibrium $\alpha'$ of $\GammaSK$ that achieves the same utility profile as $\alpha$.
\end{lemma}

\begin{proof}
Let $\alpha'$ be the probability distribution on $\times_{i\in N} A'_i$ that corresponds to
sampling an action profile $a=(a_1,\ldots , a_N)\in\times_{i\in N} A_i$ according to $\alpha$ and outputting an action profile
$a'=(\SEnc_{sk}(a_1),\dots,\SEnc_{sk}(a_N))$, where $sk$ is the secret key generated by $\SGen$.
Note that $\alpha'$ achieves the same utility profile as $\alpha$ by construction.

To show that such $\alpha'$ constitutes a correlated equilibrium of $\GammaSK$,
we need to verify that the conditions from Definition \ref{def:CE} are satisfied, i.e.
for every player $i$ and for all $b'_i,\hat{a}'_i\in A'_i$ it must hold that
\begin{equation}\label{eqn:CCE_SK_cond}
	\E_{sk\larr\SGen(),a'\larr\alpha'}[u'_i(a')|a'_i=b'_i]
		\geq \E_{sk\larr\SGen(),a'\larr\alpha'}[u'_i(\hat{a}'_i,a'_{-i})|a'_i=b'_i].
\end{equation}

Since $\SKE$ is perfectly secure, it follows from Definition~\ref{def:perfectSecurity} that for any $a'_0,a'_1\in A_i'$,
\[
\E_{sk\larr\SGen(),a'\larr\alpha'}[u'_i(a')|a'_i=a'_0]=\E_{sk\larr\SGen(),a'\larr\alpha'}[u'_i(a)|a'_i=a'_1].
\]

Thus, for any player $i$, the expected utility from the distribution $\alpha'$ is independent
of the advice $a'_i$.
Moreover, since the underlying encryption scheme is perfectly non-malleable (Definition \ref{def:perfNML}),
no player $i$ can generate (with any advantage\footnote{More precisely, no player can generate
such a deviation $a^*_i$ with more success than by random guessing.}) a deviation $a^*_i$ satisfying
$R(a^*_i,a_i)$ for any known relation $R$.
It follows that we need only to consider deviations $a^*_i$ that are independent of the received advice $a_i$.
Therefore, equation~\ref{eqn:CCE_SK_cond} can be rewritten as the following:
for every player $i$ and for all $\hat{a}'_i\in A'_i$ independent of $a'_i$,
\[
	\E_{sk\larr\SGen(),a'\larr\alpha'}[u'_i(a')]
		\geq \E_{sk\larr\SGen(),a'\larr\alpha'}[u'_i(\hat{a}'_i,a'_{-i})],
\]
which holds because $\alpha'$ is by Lemma~\ref{lem:rhoEquivalenceOfCCE} a coarse correlated equilibrium of $\GammaSK$.
\end{proof}

\Pcomment{Should say something about encrypting multiple messages with the same key. This is in general not kosher even though the scheme has perfect secrecy - the above proof goes through since each player gets only a SINGLE ciphertext!}


\begin{lemma}\label{lem:CCEvsCEinPKE}
Let $\PKE=(\PGen,\PEnc,\PDec)$ be a $\CCA$-secure public-key encryption scheme.
Then for any strategic game $\Gamma$, it holds that for any computational coarse correlated equilibrium $\alpha$ of $\Gamma$
there exists a computational correlated equilibrium $\alpha'$ of $\GammaPK$ that achieves the same utility profile as $\alpha$.
\end{lemma}
\begin{proof}
For each security parameter $k\in\NN$, let $(pk,sk)$ be the corresponding key pair generated by $\PGen(1^k)$.
Consider the following probability ensemble $\alpha'=\{\alpha^{\prime{}(k)}\}_{k\in\NN}$ on $\{\times_{j\in N} A_{j}^{\prime{}(k)}\}_{k\in\NN}$
that corresponds for each $k\in\NN$ to sampling an action profile $a=(a_1,\ldots , a_N)\in\times_{i\in N} A_i$ according to $\alpha^{(k)}$
and outputting an action profile $a'=(\PEnc_{pk}(a_1),\dots,\PEnc_{pk}(a_N))$.
Note that $\alpha'$ achieves the same utility profile as $\alpha$ by construction.

Assume that $\alpha'$ is not a computational correlated equilibrium of $\GammaPK$ (Definition \ref{def:compCE}), i.e.
there exist a player $i\in N$, a \PPT{}-samplable ensemble $\hat{\alpha}_{i}^{\prime{}}=\{\hat{\alpha}_{i}^{\prime{}(k)}\}_{k\in\NN}$
on $\{A_{i}^{\prime{}(k)}\}_{k\in\NN}$, and a non-negligible function $\delta(\cdot)$ such that for every $k\in\NN$
\begin{equation}\label{eqn:CCE_PK_cond}
	\E_{\substack{(pk,sk)\larr\PGen(1^k),\\a'\larr\alpha^{\prime{}(k)}}}
		 [u_{i}^{\prime{}(k)}(a')]
	\leq
	\E_{\substack{(pk,sk)\larr\PGen(1^k),\\a'\larr\alpha^{\prime{}(k)},\hat{a}_{i}^{\prime{}}\larr\hat{\alpha}_{i}^{\prime{}(k)}(a'_i)}}
		 [u_{i}^{\prime{}(k)}(\hat{a}_{i}^{\prime{}},a'_{-i})]-\delta(k).
\end{equation}
We show that one can use such a deviation $\hat{\alpha}_{i}^{\prime{}}$ to construct a \PPT{} adversary
that contradicts the computational non-malleability of the encryption scheme $\PKE$ (Definition \ref{def:compNM}).

Let $\Adv$ be the adversary that for each security parameter $k\in\NN$ behaves as follows.
$\Adv$ receives a public key $pk$ from the challenger and sends back $M=\alpha^{(k)}$ as the message distribution.
Upon receiving the challenge ciphertext $c$ the adversary $\Adv$ samples $c'\larr\hat{\alpha}_{i}^{\prime{}(k)}(c)$ and sends $c'$ to the challenger
together with the relation
\[
	R(b,\hat{b})=
	\begin{cases}
	1							&\text{w.p. } \frac{1}{2}\cdot(\E_{a\larr\alpha^{(k)}}[u_i(\hat{b},a_{-i})|a_i=b]-\E_{a\larr\alpha^{(k)}}[u_i(a_i,a_{-i})|a_i=b]+1),\\
	0 &\text{otherwise.}
  \end{cases}
\]
We can assume without loss of generality that all the utilities of all the players in $\Gamma$ are between $0$ and $1$
(the corresponding linear transformation of the game matrix does not change the strategic properties of the game), hence the above expression
defining the probability that $R(b,\hat{b})$ holds is between $0$ and $1$.
Note that $M$ is efficiently samplable and that the relation $R$ is efficiently computable.

Consider the success probability of $\Adv$ in the experiment $\PubK_{\Adv,\PKE}^{\NM}(k)$, i.e.
\begin{align*}
	\Pr[\PubK_{\Adv,\PKE}^{\NM}(k) = 1]
	&=
	\Pr_{\substack{(pk,sk)\larr\PGen(1^k)\\a\larr\alpha^{(k)},\hat{a}'_{i}\larr\hat{\alpha}_{i}^{\prime{}(k)}(\PEnc_{sk}(a_i))}}
		[\hat{a}'_i\neq\PEnc_{sk}(a_i) \land R(a_i,\PDec_{sk}(\hat{a}'_{i}))]\\
	{}&=
	\frac{1}{2}\bigg(\E_{\substack{(pk,sk)\larr\PGen(1^k),\\a'\larr\alpha^{\prime{}(k)},\hat{a}_{i}^{\prime{}}\larr\hat{\alpha}_{i}^{\prime{}(k)}(a'_i)}}
	[u_{i}^{\prime{}(k)}(\hat{a}_{i}^{\prime{}},a'_{-i})]
	-
	\E_{\substack{(pk,sk)\larr\PGen(1^k),\\a'\larr\alpha^{\prime{}(k)}}}
	[u_{i}^{\prime{}(k)}(a')] + 1 \bigg).
\end{align*}
Note that the scaling needed for relation $R$ is done by some finite factor, since the game matrix of $\Gamma$
does not depend on the security parameter $k$.
Therefore, it follows from equation \ref{eqn:CCE_PK_cond} that this probability is larger than $\delta'(k)$ for some non-negligible
function $\delta'(\cdot)$.

On the other hand, the success probability of $\Adv$ in the experiment $\PubK_{\Adv,\PKE}^{\NM,\$}(k)$, i.e.
\begin{align*}
	\Pr[\PubK_{\Adv,\PKE}^{\NM,\$}(k)=1]
	&=
	\Pr_{\substack{(pk,sk)\larr\PGen(1^k)\\a,\tilde{a}\larr\alpha^{(k)},\hat{a}'_{i}\larr\hat{\alpha}_{i}^{\prime{}(k)}(\PEnc_{sk}(a_i))}}
		[\hat{a}'_i\neq\PEnc_{sk}(a_i) \land R(\tilde{a}_i,\PDec_{sk}(\hat{a}'_{i}))]\\
	{}&=
	\frac{1}{2}\bigg(\E_{\substack{(pk,sk)\larr\PGen(1^k),\\a'\larr\alpha^{\prime{}(k)},\hat{a}_{i}^{\prime{}}\larr\hat{\alpha}_{i}^{\prime{}(k)}}}
	[u_{i}^{\prime{}(k)}(\hat{a}_{i}^{\prime{}},a'_{-i})]
	-
	\E_{\substack{(pk,sk)\larr\PGen(1^k),\\a'\larr\alpha^{\prime{}(k)}}}
	[u_{i}^{\prime{}(k)}(a')] + 1 \bigg),
\end{align*}
can be at most $\epsilon(k)$ for some negligible function $\epsilon$.
This follows from the fact that $\alpha$ is a computational coarse correlated equilibrium,
and no independent deviation can yield a non-negligible improvement in expectation on the utility of any player $i$.

Putting the above two observations together we conclude that for some non-negligible $\delta^{*}(\cdot)$
\[
\left|\Pr[\PubK_{\Adv,\PKE}^{\NM}(k)=1]-\Pr[\PubK_{\Adv,\PKE}^{\NM,\$}(k)=1]\right|\geq \delta^{*}(k),
\]
a contradiction to computational non-malleability of $\PKE$.
\end{proof}

\subsection{What can I do with an encrypted action?}\label{subsec:whatCanIDo}

We employ blinded games as a tool to achieve equilibria in the underlying game.
The pre-play protocols in the next section will issue ``advice'' to the players as \emph{encrypted actions},
that is, actions in the blinded game. In this section we address how an action of the blinded game
can be ``used'' to take a corresponding action in the underlying game.

We return to the concept of verifiability of mediation, introduced in Section~\ref{sec:intro}.
Since the players do not know the secret key associated with a blinded game,
they cannot decrypt an encrypted action (and indeed, this is an essential
property upon which the pre-play protocols will depend).
The players therefore invoke a third party who plays the underlying game \emph{on their behalf}.
The third party will act in a way which can be publicly
verified, so no trust need be placed in him to perform actions correctly: if he misbehaves, then the misconduct will be
detected and he can be held accountable.
This is in contrast to the usual idea of trusted mediation for implementation of equilibria.

The importance of reducing the trust placed in mediators has long been recognized in the literature,
and the first formal definition of a verifiable but not trusted form of mediation was given in \cite{ILM11}, which
introduced the concept of \emph{verifiable mediator}.

\begin{definition}[Verifiable mediator \cite{ILM11}]
A \emph{verifiable mediator} is a mediator which performs all actions in a publicly verifiable way,
and does not use any information that must be kept secret.
\end{definition}

We introduce the new concept of a \emph{verifiable proxy}, which is used in our construction.
Note that the new concept is incomparable to the verifiable mediator of \cite{ILM11}.

\begin{definition}[Verifiable proxy]
A \emph{verifiable proxy} is a mediator which performs all actions in a publicly verifiable way,
and does not give the players any information that affects their strategic choices in the underlying strategic game.
\end{definition}

In our setting, the (only) action that the verifiable proxy performs for the players is to \emph{translate}
the action from an encrypted form to the original form. It is well known that decryption can be done verifiably
(see Appendix~\ref{appx:verifiableDecryption} for details). Importantly, the verifiable proxy acts independently for each player:
the correlation between players' strategies is achieved by the players themselves with no external help, and the verifiable proxy acts
simply as a \emph{proxy} or interface so that the players may use encrypted actions to play in the underlying game.


We believe that (in contrast to general trusted mediators), verifiable proxies are a very realistic and mild requirement
in many scenarios, since many games are already ``set up'' by some entity (e.g. the stock exchange or an online games company),
which could easily set up instead a version of the game incorporating encrypted actions.
Moreover, the impossibility result of \cite{HNR13} shows that without any mediation, even correlated equilibria cannot in general
be achieved by cheap talk: so some weak notion of mediation is necessary in order to bypass this result and give useful correlated equilibrium implementations.


\begin{example}
More concretely, we provide a toy example involving the well-known ``Battle of the Sexes'' game (Figure~\ref{fig:battleSexes}),
where two friends are deciding on a joint activity, and they have opposing preferences but would rather be together than apart:

\begin{figure}[h!]
\bgroup
\def\arraystretch{1}
\begin{center}
\begin{tabular}{r|c|c|}
\multicolumn{1}{r}{}
 &  \multicolumn{1}{c}{Bach ($B$)}
 & \multicolumn{1}{c}{Stravinsky ($S$)} \\
\cline{2-3}
~~~Bach ($B$)~~~ & $\mathbf{2,5}$ & $0,0$ \\
\cline{2-3}
~~~Stravinsky ($S$)~~~ & $0,0$ & $\mathbf{5,2}$ \\
\cline{2-3}
\end{tabular}
\end{center}
\egroup
\caption{``Battle of the Sexes'' game}
\label{fig:battleSexes}
\end{figure}

It is a correlated equilibrium to randomize over $(B,B)$ and $(S,S)$. In this scenario, the ``encrypted advice''
could be an order to an online ticket vendor for either a Bach or Stravinsky concert,
encrypted under the public key of the vendor. The set-up assumption here would be that the online vendor has published a public key
and accepts encrypted orders. Since accepting orders in a variety of formats desirable to customers is in the vendor's interest,
we consider this to be a very feasible scenario.

Note that as this particular example is a correlated equilibrium, it is unnecessary to encrypt advice
(e.g. since the protocol of \cite{DHR00} applies). However, the example serves to illustrate
that verifiable translation can be a highly realistic and mild assumption.
\end{example}

\section{Our Protocols}\label{sec:protocols}
In this section we give cryptographic protocols (in the computational and information-theoretic settings) that
achieve the utility profile of any coarse correlated equilibrium.

\subsection{Cryptographic cheap talk}




\begin{definition}[Cheap talk extension \cite{DHR00,GLR10}]
For a strategic game $\Gamma$, the \emph{cheap talk extension} $\widetilde{\Gamma}$
is defined as an extensive game consisting of a pre-play phase
in which the players exchange messages, followed by the play in
the original strategic game. The communication is non-binding (unlike in signaling games)
in that it does not directly affect players' utilities in the underlying game,
that is, players' utilities in the cheap talk extension depend only on actions
taken in the strategic game.
The \emph{cryptographic cheap talk extension} is defined exactly like the cheap talk extension,
except that the players exchange messages
during a polynomially bounded number of rounds prior to the play in the
original game $\Gamma$.
\end{definition}




We follow the pre-play paradigm of \cite{barany1992}, where the mediator is replaced
by ``cheap talk'' communication prior to game play.
We construct protocols to be run during pre-play, which implement any (computational) coarse correlated equilibrium of blinded games
as a (computational) Nash equilibrium of the (computational) cheap talk extension.


\subsection{Protocol for computationally bounded players}

In this protocol, the players run a computationally secure multi-party computation to sample an action profile from any computational correlated equilibrium of the blinded game.

\begin{framed}
\begin{center}
\textbf{Protocol \refprot{prot:computational}.} Implementing any computational correlated equilibrium $\alpha'$ of $\GammaPK$:
\end{center}
Let $\PKE=(\PGen,\PEnc,\PDec)$ be a $\CCA$-secure public-key encryption scheme
and let $(pk,sk)\larr\PGen(1^k)$ with $pk$ known to all players.
Communication is via broadcast.
\begin{enumerate}
	\item The players run a computationally secure multi-party
		computation protocol (secure against $t\leq N-1$ corruptions) to implement
		the function that samples an action profile $a'\larr\alpha'$,
    		and outputs to each player $i$ his action $a'_i$.
	\item Every player takes $a'_i$ as its action in $\GammaPK$.
\end{enumerate}
\end{framed}

We show that rational computationally bounded players will follow the above protocol,
so they can use it to implement any computational correlated equilibrium.
Then, by combining the above with our results from Section~\ref{sec:blindedGames} about
correspondence of coarse correlated equilibria in the underlying game and correlated equilibria in its blinded version,
we obtain that the protocol can moreover be used to implement any computational \emph{coarse} correlated equilibrium.

Note that it is necessary to treat the two-player case somewhat differently from the case with three or more players,
because of the problem of guaranteed output delivery in the two-player case (which was described in Section~\ref{subsec:mpc}).
We begin by presenting the simpler Theorem~\ref{thm:PKprotocolThree}, which states that Protocol~\ref{prot:computational} can
be \emph{directly} used by three or more players to implement any computational coarse correlated equilibrium.
Then, we give Theorems~\ref{thm:PKprotocol} and \ref{thm:PKprotocolGeneral} which show that by running a \emph{slightly modified}
version of Protocol~\ref{prot:computational}, it is possible for \emph{any} number of players to
implement any computational coarse correlated equilibrium.

\begin{theorem}\label{thm:PKprotocolThree}
Let $\PKE=(\PGen,\PEnc,\PDec)$ be a $\CCA$-secure public-key encryption scheme,
and let $\Gamma$ be any finite strategic game with three or more players.
For any computational coarse correlated equilibrium $\alpha$ of $\Gamma$,
there exists a computational Nash equilibrium $\widetilde{\alpha}$ of the computational cheap talk
extension $\widetilde{\GammaPK}$ that achieves the same utility profile as $\alpha$.
\end{theorem}
\begin{proof}
Let $\alpha'$ be the computational correlated equilibrium of $\GammaPK$ from Lemma~\ref{lem:CCEvsCEinPKE}
that achieves the same utility profile as $\alpha$.
We show that using Protocol~\ref{prot:computational} in order to implement $\alpha'$ constitutes a computational Nash equilibrium
in the cryptographic cheap talk extension $\widetilde{\GammaPK}$. Note that it is payoff-equivalent to $\alpha$ by construction.

By the \emph{privacy} guarantee of the secure multi-party computation protocol, we have that no player can learn any 
(non-negligible amount of) information that cannot be deduced from his intended output in the first place, 
even if he deviates from the protocol arbitrarily. Moreover,
since there are three or more players and we consider only unilateral\footnote{That is, we only consider deviations
from the protocol by a single (malicious) player, rather than by coalitions of multiple colluding players.} 
deviations (as implied by the definition of Nash equilibrium), the protocol has the property of
\emph{guaranteed output delivery}\footnote{We remark that in fact, the slightly weaker property of \emph{fairness} is sufficient: that is,
the property that if any player receives his output in the protocol, then every honest player will receive her correct output too.
However, in the settings we consider, the stronger property of \emph{guaranteed output delivery} is known to hold, 
hence we refer to the latter property in order to slightly simplify the proof.}: 
therefore, the deviation of any player $i$ cannot prevent any other player $j$ from receiving her correct output $a'_j$.

We have shown that for any player, there is no deviation during the protocol phase that is profitable by more than negligible amount.
Hence, we consider only the case where each player $i$ receives his correct output $a'_i$.
Since $\alpha'$ is, by Lemma~\ref{lem:CCEvsCEinPKE}, a computational correlated equilibrium of $\GammaPK$,
no player has an incentive to deviate from the prescribed advice,
and thus the players will play according to the sampled action profile $a'$.
Therefore, to follow Protocol~\ref{prot:computational} is the computational Nash equilibrium $\widetilde{\alpha}$
of $\widetilde{\GammaPK}$ payoff-equivalent to $\alpha$.
\end{proof}

\subsubsection{Dealing with the two-player case}\label{sec:twoPlayer}

In the two-player case, the additional complication stems from the fact that
in this setting we do not have guaranteed output delivery:
hence, it is necessary to consider that a player may be incentivized to cause a protocol execution to terminate prematurely.
In order to disincentivize such behavior, we introduce an additional ``punishment'' condition to the protocol, as follows.

\begin{framed}
\begin{center}
\textbf{Protocol \refprot{prot:compWithNE}.} Implementing any computational correlated equilibrium $\alpha'$ of $\GammaPK$:
\end{center}
Let $\PKE=(\PGen,\PEnc,\PDec)$ be a $\CCA$-secure public-key encryption scheme
and let $(pk,sk)\larr\PGen(1^k)$ with $pk$ known to all players.
Communication is via broadcast.
\begin{itemize}
	\item The players run Protocol \ref{prot:computational} as long as no player is detected to deviate from the protocol.
	\item If any player $i$ is detected to deviate from the protocol, then all (other) players adopt the strategies (in $\GammaPK$) corresponding to the worst Nash equilibrium $\sigma^{i}$ for player $i$.
\end{itemize}
\end{framed}

Using Protocol~\ref{prot:compWithNE}, we obtain the following theorem that applies for \emph{any} number of players.

\begin{theorem}\label{thm:PKprotocol}
Let $\PKE=(\PGen,\PEnc,\PDec)$ be a $\CCA$-secure public-key encryption scheme,
and let $\Gamma$ be any finite strategic game.
For any computational coarse correlated equilibrium $\alpha$ of $\Gamma$ that for each player
achieves at least as high utility as the worst Nash equilibrium,
there exists a computational Nash equilibrium $\widetilde{\alpha}$ of the computational cheap talk
extension $\widetilde{\GammaPK}$ that achieves the same utility profile as $\alpha$.
\end{theorem}
\begin{proof}[Proof]
Let $\alpha'$ be the computational correlated equilibrium of $\GammaPK$ from Lemma~\ref{lem:CCEvsCEinPKE}
that achieves the same utility profile as $\alpha$.
We show that using Protocol~\ref{prot:compWithNE} in order to implement $\alpha'$ constitutes a computational Nash equilibrium
in the cryptographic cheap talk extension $\widetilde{\GammaPK}$. 
For any security parameter $k$, the following events may occur
during the run of the protocol:
\begin{enumerate}
\item \label{itm:learnsEarly} a player learns its advice before the other players;
\item \label{itm:deviatesNoticed} a player deviates from the protocol and the deviation is detected by the other players; or
\item \label{itm:deviatesUnnoticed} a player deviates from the protocol and it is unnoticed.
\end{enumerate}

Addressing (\ref{itm:learnsEarly}): it follows from $\CCA{}$-security of the public-key encryption scheme $\PKE$
(Definition~\ref{def:CCAsec}) that each player is indifferent (up to a negligible improvement in utility)
between any advice he may receive, and thus gains no advantage from learning
his advice first. In particular, he has no incentive to abort the protocol
and prevent others from learning their advice.
Addressing (\ref{itm:deviatesNoticed}): the expectation of any player $i$ in the default Nash equilibrium
$\sigma^i$ is at most the expectation of player $i$ in $\alpha$.
Addressing (\ref{itm:deviatesUnnoticed}): the security of the multi-party computation protocol ensures that players can cheat
without being caught with at most negligible probability.
Thus, the increase in utility from any cheating strategy is at most negligible.

There is no deviation during the protocol phase profitable by more than negligible amount.
Consider the case that every player $i$ received his advice $a'_i$. Since $\alpha'$
is, by Lemma~\ref{lem:CCEvsCEinPKE}, a computational correlated equilibrium of $\GammaPK$,
no player has an incentive to deviate from the prescribed advice,
and the players will play according to the sampled action profile $a'$.
Therefore, to follow Protocol~\ref{prot:compWithNE} is the computational Nash equilibrium $\widetilde{\alpha}$
of $\widetilde{\GammaPK}$ payoff-equivalent to $\alpha$.
\end{proof}

It is possible to eliminate the condition (from Theorem~\ref{thm:PKprotocol}) that the implemented coarse correlated equilibrium
does at least as well as the respective worst Nash equilibrium for each player, thereby obtaining a yet more general theorem
as follows.

\begin{theorem}\label{thm:PKprotocolGeneral}
Let $\PKE=(\PGen,\PEnc,\PDec)$ be a $\CCA$-secure public-key encryption scheme,
and let $\Gamma$ be any finite strategic game.
For any coarse correlated equilibrium $\alpha$ of $\Gamma$,
there exists a computational Nash equilibrium $\widetilde{\alpha}$ of the computational cheap talk
extension $\widetilde{\GammaPK}$ that achieves the same utility profile as $\alpha$.
\end{theorem}

The proof of Theorem~\ref{thm:PKprotocolGeneral} makes use of another variant of Protocol~\ref{prot:computational}.
The details of this variant protocol (Protocol~\ref{prot:compWithMinMax}) are given in Appendix~\ref{appx:minMax}
along with the proof of the theorem.

We remark that Protocol~\ref{prot:compWithNE} has certain more desirable properties than Protocol~\ref{prot:compWithMinMax}:
in particular, Protocol~\ref{prot:compWithNE} is \emph{free of empty threats}, which ensures that Nash equilibria
in the protocol are stable even when players may change strategy \emph{adaptively} during protocol execution
(a formal definition of empty threats may be found in Appendix~\ref{appx:emptyThreats}).
Ultimately, notwithstanding the restriction on the class of achieved coarse correlated equilibria,
we consider Theorem~\ref{thm:PKprotocol} to be the much stronger result compared to Theorem~\ref{thm:PKprotocolGeneral},
for the following reasons:
\begin{itemize}
\item all coarse correlated equilibria that players might rationally wish
to implement by cheap talk do dominate all Nash equilibria (otherwise, they could achieve a better payoff from a Nash equilibrium
without the hassle of a pre-play protocol); and
\item unlike Protocol~\ref{prot:compWithMinMax}, Protocol~\ref{prot:compWithNE} is free of empty threats; and
\item the expected payoff even when the protocol is aborted and the default strategy invoked is higher in Protocol~\ref{prot:compWithNE} than in Protocol~\ref{prot:compWithMinMax}.
\end{itemize}

\paragraph{Strategic equivalence}
Lemma~\ref{lem:PKstrategicEquivalence}, below, proves the strategic equivalence of the
cryptographic cheap talk extension $\widetilde{\GammaPK}$ to the underlying game $\Gamma$.

\begin{lemma}\label{lem:PKstrategicEquivalence}
Let $\PKE=(\PGen,\PEnc,\PDec)$ be a $\CCA$-secure public-key encryption scheme,
and let $\Gamma$ be any finite strategic game.
For any computational Nash equilibrium $\widetilde{\alpha}$ of the cryptographic cheap talk extension $\widetilde{\GammaPK}$,
there exists a computational coarse correlated equilibrium $\alpha$ of $\Gamma$ that achieves the same utility profile as $\widetilde{\alpha}$.
\end{lemma}
\begin{proof}
We show that the probability ensemble $\alpha$ induced by $\widetilde{\alpha}$ on action profiles of $\Gamma$
is a computational coarse correlated equilibrium of $\Gamma$.

Assume that $\alpha$ is not a computational coarse correlated equilibrium, i.e. there exists a player $i$ that has a \PPT{}-samplable unilateral deviation to
$\alpha$ that improves his expectation for every $k\in\NN$ by $\delta(k)$ for some non-negligible $\delta(\cdot)$.
However, such deviation can be used by player $i$ also against $\widetilde{\alpha}$ to gain a non-negligible improvement in his expectation in $\widetilde{\GammaPK}$,
a contradiction to the fact that $\widetilde{\alpha}$ is a computational Nash equilibrium of $\widetilde{\GammaPK}$.
\end{proof}

\begin{corollary}
For any finite strategic game $\Gamma$, the cryptographic cheap talk extension $\widetilde{\GammaPK}$ is
strategically equivalent to $\Gamma$, that is, for every Nash equilibrium $\widetilde{\alpha}$ of $\widetilde{\GammaSK}$, there exists
a coarse correlated equilibrium of $\Gamma$ that achieves the same utility profile as $\widetilde{\alpha}$, and vice versa.
\end{corollary}
\begin{proof}
Follows immediately from Lemma~\ref{lem:PKstrategicEquivalence} and Theorem~\ref{thm:PKprotocol} (or Theorem~\ref{thm:PKprotocolThree}
for the case of three or more players).
\end{proof}

\subsection{Protocol for computationally unbounded players}

An alternative protocol using secret-key encryption implements all coarse correlated equilibria --
not just computational ones -- for all strategic games with four or more players.
As discussed in Section~\ref{sec:intro}, the condition of four or more players is unavoidable.
In this (more traditional) setting, the players are computationally unbounded.

\vspace{2em}
\begin{framed}
\begin{center}
\textbf{Protocol \refprot{prot:perfect}.} Implementing any correlated equilibrium $\alpha'$ of $\GammaSK$:
\end{center}
Let $\SKE=(\SGen,\SEnc,\SDec)$ be a perfectly non-malleable and verifiably decryptable secret-key encryption scheme
and let $sk\larr\SGen$.
Let each player $i$ possess a distinct share $sk_i$ of an $(N-1)$-out-of-$N$ secret-sharing $\{sk_1,\dots,sk_N\}$ of $sk$.
Communication is via pairwise channels.
\begin{enumerate}
	\item The players run a perfectly secure multi-party
		computation to implement the function that
		samples a profile $a'\larr\alpha'$,
    		and outputs to each $i$ his action $a'_i$.
	\item Every player takes $a'_i$ as its action in $\GammaSK$.
\end{enumerate}
\end{framed}

\begin{theorem}\label{thm:SKprotocol}
Let $\SKE=(\SGen,\SEnc,\SDec)$ be a perfectly non-malleable and verifiably decryptable secret-key encryption scheme,
and let $\Gamma$ be any finite strategic game with four or more players.
For any coarse correlated equilibrium $\alpha$ of $\Gamma$ there exists a Nash equilibrium $\widetilde{\alpha}$ of the cheap talk
extension $\widetilde{\GammaSK}$ that achieves the same utility profile as $\alpha$.
\end{theorem}
\begin{proof}
Let $\alpha'$ be the correlated equilibrium of $\GammaSK$ from Lemma~\ref{lem:CCEvsCEinSKE} that achieves the same utility profile as $\alpha$. We show that to follow Protocol~\ref{prot:perfect} in order to implement $\alpha'$ constitutes the Nash equilibrium $\widetilde{\alpha}$ in the cryptographic cheap talk extension $\widetilde{\GammaSK}$ that achieves the same utility profile as $\alpha$.

First note that since the players are using a perfectly secure protocol with output guarantee (see Section~\ref{subsec:mpc}) to implement sampling from $\alpha'$, no player can prevent the others from learning their advice by a unilateral deviation during the multi-party computation phase. Moreover, even if a single player $i$ withholds its share $sk_i$ the remaining players hold $N-1$ shares of the secret key $sk$ that are sufficient to reconstruct the secret key and sample an action profile from $\alpha'$. Hence, any unilateral deviation does not influence the distribution on actions taken by the other players. Assume that there exists a unilateral deviation for some player $i$ in $\widetilde{\GammaSK}$ that allows him to gain a higher utility than by playing according to $\widetilde{\alpha}$. This contradicts $\alpha'$ being a correlated equilibrium of $\GammaSK$, since it could be used as a unilateral profitable deviation against $\alpha'$ in $\GammaSK$ as well.
\end{proof}

\paragraph{Strategic equivalence}
Lemma~\ref{lem:SKstrategicEquivalence}, below, proves the strategic equivalence of the
cheap talk extension $\widetilde{\GammaSK}$ to the underlying game $\Gamma$.
\begin{lemma}\label{lem:SKstrategicEquivalence}
Let $\SKE=(\SGen,\SEnc,\SDec)$ be a perfectly non-malleable and verifiably decryptable secret-key encryption scheme,
and let $\Gamma$ be any finite strategic game with four or more players.
For any Nash equilibrium $\widetilde{\alpha}$ of the cheap talk extension $\widetilde{\GammaSK}$,
there exists a coarse correlated equilibrium $\alpha$ of $\Gamma$ that achieves the same utility profile as $\widetilde{\alpha}$.
\end{lemma}
\begin{proof}
We show that the distribution $\alpha$ induced by $\widetilde{\alpha}$
on action profiles of $\Gamma$ is a coarse correlated equilibrium
of $\Gamma$.
Suppose $\alpha$ is not a coarse correlated equilibrium, i.e. there exists a player $i$ that has a deviation to $\alpha$ which improves his expectation.
However, such a deviation contradicts the fact that $\widetilde{\alpha}$ is a Nash equilibrium of $\widetilde{\GammaSK}$,
since it is also a profitable unilateral deviation against $\widetilde{\alpha}$ in $\widetilde{\GammaSK}$.
\end{proof}

\begin{corollary}
For any game $\Gamma$, it holds that the cheap talk extension $\widetilde{\GammaSK}$ is
strategically equivalent to $\Gamma$, that is, for every Nash equilibrium $\widetilde{\alpha}$ of $\widetilde{\GammaSK}$, there exists
a coarse correlated equilibrium of $\Gamma$ that achieves the same utility profile as $\widetilde{\alpha}$, and vice versa.
\end{corollary}
\begin{proof}
Follows immediately from Theorem~\ref{thm:SKprotocol} and Lemma~\ref{lem:SKstrategicEquivalence}.
\end{proof}

\paragraph{Sequential equilibrium}
We also show that the equilibrium from Theorem~\ref{thm:SKprotocol} is a \emph{sequential equilibrium}
(relevant formal definitions are given in Appendix~\ref{appx:extensiveGames}):
informally, we show that by following the prescribed strategy, the players are making optimal decisions at all points in the game tree.
Our proof relies on perfect security for multi-party computation protocols in the presence of one actively corrupted  and one passively corrupted party which can be achieved only for six or more players (as shown by Fitzi, Hirt and Maurer~\cite{DBLP:conf/crypto/FitziHM98}, see Section~\ref{subsec:mpc}). Hence, the statement of the following theorem is less general than the statement of Theorem~\ref{thm:SKprotocol}.

\begin{theorem}
Let $\SKE=(\SGen,\SEnc,\SDec)$ be a perfectly non-malleable and verifiably decryptable secret-key encryption scheme,
and let $\Gamma$ be any finite strategic game with six or more players.
For any coarse correlated equilibrium $\alpha$ of $\Gamma$ there exists a sequential equilibrium $(\widetilde{\alpha},\mu)$ of the cheap talk
extension $\widetilde{\GammaSK}$ that achieves the same utility profile as $\alpha$.
\end{theorem}
\begin{proof}
We assume without loss of generality that the multi-party computation protocol has the canonical structure where at each round a single player receives a message from one of the other players (i.e. the information sets in the extensive game correspond to histories consistent with the received message). Since there is at least six players, we can assume that multi-party computation is secure in the presence of one static and one active corruption.
Consider the behavioral strategy profile $\widetilde{\alpha}$ corresponding to following Protocol~\ref{prot:perfect} at each history where a player receives a message from some other player (in particular this corresponds to ignoring all received messages after termination of the multi-party computation).

First, we specify the belief system $\mu$ of players at any information set. The beliefs at
any information set on the equilibrium path are derived from the behavioral strategy $\widetilde{\alpha}$ by Bayes' rule, and for any information set $I$
that lies off the equilibrium path (i.e. an information set corresponding to receiving a message out of the scope of the protocol), let $\mu(I)$ be the uniform distribution on all histories in $I$.
To show that $(\widetilde{\alpha},\mu)$ is a sequential equilibrium, we must show that $(\widetilde{\alpha},\mu)$ is both sequentially rational and
consistent.

Since $\widetilde{\alpha}$ is a Nash equilibrium (as shown in Theorem~\ref{thm:SKprotocol}), the behavioral strategy to follow $\widetilde{\alpha}$ is optimal for any information set on the equilibrium path.
Hence, to conclude that $(\widetilde{\alpha},\mu)$ is sequentially rational, we just need to show that $\widetilde{\alpha}$ is also optimal off the equilibrium path, given the beliefs of $\mu$. Let $I$ be an information set of player $i$ at some point off the equilibrium path that corresponds to receiving a message from player $j$. Note that even if $j$ sends to $i$ its complete view of the protocol up to this point player $i$ cannot use such information to produce a profitable deviation, since such deviation would imply an adversary corrupting actively player $i$ and statically player $j$ able to break the perfect security of the multi-party computation protocol. Now consider any history off the equilibrium path after the termination of the multi-party computation, and assume that player $i$ receives the private advice of some other player. There cannot exist a profitable deviation of player $i$, since such a deviation would contradict security of the secret key encryption scheme.

To show that $(\beta,\mu)$ is consistent we use the ``trembling-hand'' approach. Consider the sequence of assessments $\{(\beta^{(n)},\mu^{(n)})\}_{n=1}^{\infty}$  where each $\beta^{(n)}$
assigns non-zero probability $\epsilon^{(n)}$ to all actions that are taken with zero probability in $\beta$, such that $\epsilon^{(n)}$ goes to zero as $n\rarr\infty$,
and the belief system $\mu^{(n)}$ is derived from $\beta^{(n)}$ using Bayes' rule.
First note that the sequence $\{(\beta^{(n)},\mu^{(n)})\}_{n=1}^{\infty}$ converges to $(\beta,\mu)$.
The sequence of behavioral strategy profiles $\{\beta^{(n)}\}_{n=1}^{\infty}$ converges to $\beta$ by construction.
Since $\mu^{(n)}$ is derived from $\beta^{(n)}$ by the Bayes' rule,
$\mu^{(n)}$ converges to $\mu$ for every information set on the equilibrium path. For every information set $I$ off the equilibrium path, the distribution $\mu^{(n)}(I)$ is equal to $\mu(I)$.
Finally, $\beta^{(n)}$ is completely mixed for all $n$, hence $(\beta,\mu)$ is consistent.
\end{proof}

\subsection{Remarks on efficiency of multi-party computation}

\paragraph{Computational setting}
With recent advances in efficiency, computationally secure multi-party
computation protocols are now being considered for practical use in various settings.
Its first large-scale deployment was to compute market clearing prices for Danish sugar beet contracts in 2008 \cite{sugarbeet}.
Subsequent advances include \cite{IPS09,DO10}.
Indeed, numerous multi-party computation implementations are
available online, such as VIFF (\texttt{viff.dk}) \cite{DGKN09}.

In the common ``pre-processing model'', where pre-processing time is available
before the main computation, yet faster protocols are possible: \cite{SPDZ} achieves secure 3-party
64-bit multiplication in 0.05 ms. This could be a very reasonable
model when the same $N$ players play multiple or repeated games.

\vskip12pt
We note that there has been a line of work starting with \cite{DHR00}, on designing multi-party computation
protocols specifically for sampling from correlated equilibrium distributions.
However, these address the two-party setting, and have not taken into account the most recent advances in
general multi-party computation techniques, so we do not consider them to be of great relevance here.

\paragraph{Perfect setting}
In the perfect setting, known protocols are less efficient;
and perfectly secure encryption is relatively inefficient
due to inherently large key sizes. Nonetheless, substantial progress has been made: the best known protocol \cite{BH08}
achieves $O(N)$ communication complexity per multiplication\footnote{The circuit that the parties want to compute is usually represented
as addition and multiplication gates, and the multiplication gates have been found to be the bottleneck for multi-party computation.},
improving on previous protocols by $\Omega(N^2)$.

We consider our information-theoretic results to be of interest primarily as
proofs of possibility, and a novel application of cryptographic techniques to game theory without computational restrictions.
Certainly, for efficiency in practice and strength of results, our computational protocols are the ones of interest.

\section{Conclusion}\label{sec:conclusion}

In this work we use standard cryptographic tools -- namely, encryption schemes --
to introduce the concept of blinded games: strategic games in which players take encrypted actions,
and in particular have the possibility to take actions they know nothing about.
Moreover, we provide cryptographic protocols that enable the players to not
rely on trusted mediators in order to achieve equilibrium payoffs.

Our approach suggest new interesting uses of cryptographic methods in game theory.
We show that our blinded games offer a viable and appealing alternative to solution
concepts based on commitment, and a particularly promising direction for
future work is to apply the paradigm of leveraging players' lack of knowledge in
order to avoid commitment, in broader settings.

\paragraph{Acknowledgements}
We are grateful to Alessandra Scafuro for raising the question of encrypting advice,
to Silvio Micali for very helpful advice on exposition, and to Jesper Buus Nielsen for detailed technical comments on the final versions.

Pavel Hub\'{a}\v{c}ek acknowledges support from the European Research Commission Starting Grant 279447; 
from the Danish National Research Foundation
and The National Science Foundation of China (grant 61061130540) for the Sino-Danish Center for the Theory of Interactive Computation,
within part of this work was performed; and from the CFEM research center, supported by the Danish Strategic Research Council.

\printbibliography

\appendix
\section*{Appendix}
\section{Extensive Games}\label{appx:extensiveGames}

Here we recall the standard definition of extensive games.

\begin{definition}[Extensive game]
An extensive game $\Gamma=\langle N,H,P,A,\mathcal{I},(u_i)\rangle$ is defined by:
\begin{itemize}
\item a finite set $N$ of players,
\item a set $H$ of all possible history sequences (with the subset of all terminal histories denoted by $Z$),
\item a player function $P:H\setminus Z \rightarrow N$
	that assigns a player to every non-terminal history,
\item a function $A$ that assigns to every non-terminal history $h\in H\setminus Z$
	a finite set of actions $A(h)=\{a:(h,a)\in H\}$ available to player $P(h)$ at $h$,
\item for each player $i\in N$, a partition $\mathcal{I}_i$ of $\{h\in H: P(h)=i\}$
	such that $A(h)=A(h')$ whenever $h$ and $h'$ are in the same $I_i\in\mathcal{I}_i$,
\item for each player $i\in N$, a utility function $u_i: Z\rightarrow \mathbb{R}$,
\end{itemize}
\end{definition}

If the partition $\mathcal{I}_i$ is trivial and each $I_i\in\mathcal{I}_i$
contains a single history for every player $i$ then we say that
the extensive game is with \emph{perfect information} (i.e., every player is perfectly informed of
all actions taken by every other player).
A strategy profile $\sigma$ of an extensive game $\Gamma$ with perfect information specifies the actions
of every player at every history, i.e., for every $h\in H$ it specifies
a probability distribution on $A(h)$ for player $i=P(h)$.

The solution concept relevant to this work in the context of
extensive games with perfect information is Nash equilibrium.

\begin{definition}[Nash equilibrium of extensive game]
\label{def:PerfectExtensiveNE}
Let $\Gamma=\langle N,H,P,A,(u_i)\rangle$ be an extensive game with perfect information.
We say that strategy profile $\sigma$ is a \emph{Nash equilibrium of $\Gamma$} if for
every player $i\in N$ and for every strategy $\sigma'_i$ of player $i$:
  \[
   \E[u_i(\sigma)]\geq \E[u_i(\sigma'_i,\sigma_{-i})],
  \]
where the expectations are taken over terminal histories sampled from the
corresponding strategy profile.
\end{definition}

%
%

\begin{definition}[Computational Nash equilibrium of extensive game]
A \emph{computational Nash equilibrium of extensive game $\Gamma=\langle N,H,P,A,(u_i)\rangle$}
is a \PPT-samplable family of strategy profiles $\{\sigma^{(k)}\}_{k\in\NN}$ for $\Gamma$ if
for every player $i\in N$ and for every \PPT-samplable strategy $\sigma'_i$ of player $i$
it holds for all large enough $k$ that
  \[
   \E[u_i(\sigma^{(k)}_i)]\geq \E[u_i(\sigma'_i,\sigma_{-i}^{(k)})] -\eps(k),
  \]
where the expectations are taken over terminal histories sampled from the
corresponding strategy profile, and $\eps$ is a negligible function.
\end{definition}

In extensive games with \emph{imperfect information}, the players are not informed about all the
actions taken by their opponents. A \emph{profile of behavioral strategies} specifies a probability distributions
on actions available to every player $i\in N$ at every information set $I_i\in\mathcal{I}_i$.
The solution concept of \emph{Nash equilibrium in behavioral strategies} is defined similarly to Definition~\ref{def:PerfectExtensiveNE}.

When reasoning about Nash equilibrium in games with imperfect information we need to take into account
also the beliefs of players about the past play at any information set.
This gives rise to the notion of assessment.

\begin{definition}[Assessment]
An \emph{assessment} in an extensive game is a pair $(\beta,\mu)$,
where $\beta$ is a profile of behavioral strategies and $\mu$ is a funcion that
assigns to every information set a probability  distribution on the histories in the information set.
\end{definition}

The following solution concept aims to circumvent the instability of Nash equilibrium of
extensive games with imperfect information.

\begin{definition}[Sequential equilibrium]
Let $\Gamma=\langle N,H,P,A,\mathcal{I},(u_i)\rangle$ be an extensive game.
The assessment $(\beta,\mu)$ is a \emph{sequential equilibrium} if it is
\begin{enumerate}
	\item \emph{sequentially rational} - for every $i\in N$, for every information set $I_i\in\mathcal{I}_i$,
		and every $\beta'_i$
		\[
			\E[u_i(\beta,\mu)|I_i]\geq\E[u_i((\beta_{-i},\beta'_i),\mu)|I_i].
		\]
	\item \emph{consistent} - there exists a sequence $\{(\beta^{(n)},\mu^{(n)})\}_{n=1}^{\infty}$ of assessments
		that converges to $(\beta,\mu)$, $\beta^{(n)}$ is completely mixed for all $n\in\NN$, and $\mu^{(n)}$ is derived from
		$\beta^{(n)}$ by Bayes' rule.
\end{enumerate}
\end{definition}

\section{Verifiable Decryption}\label{appx:verifiableDecryption}

The following is the standard procedure for a \emph{prover} to convince a \emph{verifier} that he has correctly decrypted a ciphertext.
Upon decrypting, the prover obtains the un-encrypted action and the randomness that was used during encryption, and presents the
verifier with these two items. Then the verifier can run the encryption algorithm for herself, and check that the resulting ciphertexts
are the same as the ones that they submitted for decryption. By the security of the encryption scheme, it would be (computationally)
infeasible for the prover to come up with $(\mathrm{decryption}, \mathrm{randomness})$ pairs that pass this check, except by running
the decryption algorithm with the correct secret key. Hence, the verifier may be assured that the prover has decrypted correctly.

Note that this verifiable decryption procedure requires that the encryption scheme, in addition to being secure\footnote{CPA security suffices.},
has the following property, which is very common among existing schemes:
\begin{itemize}
\item \emph{Recoverable randomness.} By running the decryption algorithm on a ciphertext $c=\Enc(m)$
with a correct secret key, the decryptor must be able to recover the randomness used for encryption.
More precisely, for any given keypair $(pk,sk)$, we require that a decryptor possessing a correct secret key can efficiently
compute \emph{some} randomness $r$ that, when inputted along with the correct message to the encryption algorithm,
outputs the ciphertext in question, i.e. $\Enc_{pk}(m;r)=c$; and moreover, the decryptor cannot (with non-negligible probability) compute
a randomness that, when inputted along with an incorrect message $m'$ to the encryption algorithm, outputs the ciphertext in question,
i.e. it is infeasible to find $r'$ such that $\Enc_{pk}(m';r')=c$.
\end{itemize}

We say that encryption schemes satisfying this property are \emph{verifiably decryptable}.

\section{Implementation of \emph{Any} CCE with Any Number of Players}\label{appx:minMax}

\begin{framed}
\begin{center}
\textbf{Protocol \refprot{prot:compWithMinMax}.} Implementing any computational correlated equilibrium $\alpha'$ of $\GammaPK$:
\end{center}
Let $\PKE=(\PGen,\PEnc,\PDec)$ be a $\CCA$-secure public-key encryption scheme
and let $(pk,sk)\larr\PGen(1^k)$ with $pk$ known to all players.
Communication is via broadcast.
\begin{itemize}
	\item The players run Protocol \ref{prot:computational} as long as no player is detected to deviate from the protocol.
	\item If any player $i$ is detected to deviate from the protocol, then all (other) players adopt joint min-max strategy (in $\GammaPK$) with the worst possible outcome for player $i$.
\end{itemize}
\end{framed}

\paragraph{Theorem \ref{thm:PKprotocolGeneral}}
Let $\PKE=(\PGen,\PEnc,\PDec)$ be a $\CCA$-secure public-key encryption scheme,
and let $\Gamma$ be any finite strategic game.
For any coarse correlated equilibrium $\alpha$ of $\Gamma$,
there exists a computational Nash equilibrium $\widetilde{\alpha}$ of the computational cheap talk
extension $\widetilde{\GammaPK}$ that achieves the same utility profile as $\alpha$.
\begin{proof}[Proof (sketch)]
The players run Protocol~\ref{prot:compWithMinMax}, in which -- by construction --
adhering to the protocol yields at least as much utility as deviation, for any given player.
The full proof follows exactly the same structure as the proof of Theorem 1 of \cite{DHR00},
and we refer the reader to their paper for details.
\end{proof}

As discussed briefly in Section~\ref{sec:twoPlayer}, implementation via Protocol~\ref{prot:compWithMinMax}
has the disadvantage (compared to Protocol~\ref{prot:compWithNE}) of \emph{empty threats}:
these are the subject of Appendix~\ref{appx:emptyThreats}.

\section{Protocol~\ref{prot:compWithNE} is Free of Empty Threats}\label{appx:emptyThreats}

We start with an informal definition and discussion of empty threats,
then proceed to a full definition and proof of empty-threat-freeness of Protocol~\ref{prot:compWithNE}.

\begin{definition}[Empty threat (informal)]\label{def:emptyThreatInformal}
An empty threat posed by a player in an extensive game is a strategy of the threatening
player at a history off the equilibrium path which is not rational from his perspective.
A threatened player can demonstrate the existence of such an empty threat by taking
a beneficial deviation that would make the threatening player refrain from following through with the
announced threat.
\end{definition}

A consequence of empty threats in a Nash equilibrium is that a strategy profile containing empty threats
is not sequentially stable, that is, players that adapt their strategies during the game would not
follow such a strategy profile. One approach to avoid empty threats is to require subgame perfect equilibrium.
However, as addressed by \cite{GLR10}, there is no obvious way to define subgame perfection in the computational setting.
Therefore, we use the computational solution concept of \cite{GLR10} and show that
Protocol~\ref{prot:compWithNE} is free of empty threats.



Now, we give the definition of empty-threat-free Nash equilibrium in extensive games 
(cf. \cite{GLR10} for details and computational version).
The definition uses the following notion of set of continuations of a strategy profile
at a given history. For a history $h\in H$, a strategy $\sigma$,
and a distribution $\tau = \tau(h)$ on $A(h)$, let
\[
\Cont(h,\sigma,\tau)=
\{\pi:(\pi\text{ differs from }\sigma\text{ only on the subgame }h)\&(\pi(h) = \tau(h))\}.
\]

\begin{definition}[Empty threat]\label{def:emptyThreatFull}
Let $\Gamma=\langle N,H,P,A,(u)_i\rangle$ be an extensive game,
and let $\sigma$ be a strategy profile.
Then:
\begin{itemize}
	\item For any history $h\in Z$, no player faces an empty threat at $h$ with respect to $\sigma$.
	\item Player $i$ \emph{faces an empty threat at history $h\in H\setminus Z$ with respect to $\sigma$}
		if $i = P(h)$ and there exists a distribution $\tau = \tau(h)$ over $A(h)$ that satisfies the
		following: for all $\pi\in\Cont(h,\sigma,\tau)$ and $\pi'\in\Cont(h,\sigma,\sigma)$
		for which no player faces an empty threat at any $h'\in H$ below $h$, it holds that
		\[ \E[u_i(\pi)]>\E[u_i(\pi')].\]
\end{itemize}
A strategy profile $\sigma$ is \emph{empty-threat-free on $h$} if for all $h'\neq\emptyset$
satisfying $(h,h')\in H$ no player faces an empty threat at $(h,h')$ with respect to $\sigma$.
\end{definition}

\begin{definition}[Empty-threat-free Nash equilibrium]\label{def:emptyThreatFreeNEFull}
Let $\Gamma=\langle N,H,P,A,(u)_i\rangle$ be an extensive game.
Strategy profile $\sigma$ is an \emph{empty-threat-free Nash equilibrium} if:
\begin{itemize}
	\item $\sigma$ is a Nash equilibrium of $\Gamma$, and
	\item for any $h \in H\setminus Z$, player $P(h)$ does not face an empty threat
		at $h$ with respect to $\sigma$.
\end{itemize}
\end{definition}

\begin{remark}The above definitions readily apply to extensive games with $n>2$ players,
even though they were originally intended for games with two players.
As noted by~\cite{GLR10}, a potential shortcoming of applying this definition in games with
more than two players is that it does not take into account collusions between players.
However, we accept that the definition addresses only unilateral deviations as is standard
in Nash equilibrium.
\end{remark}	

\begin{theorem}
{Let $\widetilde{\alpha}$ be the computational Nash equilibrium
of $\widetilde{\GammaPK}$ from Theorem~\ref{thm:PKprotocol}.
Then $\widetilde{\alpha}$ is an empty-threat-free computational
Nash equilibrium of $\widetilde{\GammaPK}$.}
\end{theorem}
\begin{proof}[Proof (sketch)]
Recall that the computational Nash equilibrium $\widetilde{\alpha}$ is
payoff-equivalent to some computational coarse correlated equilibrium $\alpha$ of $\Gamma$
(Theorem~\ref{thm:PKprotocol}),
and that $\alpha$ achieves at least as high utility for each player $i$ as his worst Nash equilibrium
$\sigma^i$ of $\Gamma$ to which the players default in case any deviation of player $i$ is detected during the protocol phase.

For every security parameter $k$, we need to show that at any history no
player is facing an empty threat
(see Definition~\ref{def:emptyThreatFull}), i.e.
we need to show that there is no history $h$ with a deviation $\tau$
such that every empty-threat-free continuation of $\tau$ improves over every
empty-threat-free continuation of $\widetilde{\alpha}$ at $h$ by more than negligible amount.

It follows from Lemma~\ref{lem:CCEvsCEinPKE} that the expectation of any player after
receiving the encrypted advice is the same as the expectation of playing $\widetilde{\alpha}$
without knowing the advice. Thus, the expectation from following the protocol is the
same at any history of the cheap talk extension.

We will use the following claim that follows immediately from $\widetilde{\alpha}$
being a computational Nash equilibrium.

\begin{claim}\label{clm:protocolGain}
Any deviation during the protocol phase that goes unnoticed can give the player
at most negligible advantage.
\end{claim}

Since any observed deviation corresponds to a history in which players default to
the Nash equilibrium $\sigma$,  no player is facing an empty threat
with respect to $\widetilde{\alpha}$ at such histories, since $\sigma$
is a Nash equilibrium.

By the definition of empty threat (Definition~\ref{def:emptyThreatFull}),
no player is facing an empty threat at the final round where players take simultaneous actions
in the strategic game, and in particular it is an empty-threat-free strategy
to play according to the received advice $a'_i$ at the terminal history.
By Claim~\ref{clm:protocolGain} any unobserved deviation in the protocol phase can yield at most negligible improvement in player's
utility, thus we get by induction that to follow $\widetilde{\alpha}$ is an empty-threat-free
continuation at any history.

Finally, Claim~\ref{clm:protocolGain} also gives that no continuation
(in particular no empty-threat-free continuation) of any deviation
at any history $h$ can improve by more than negligible amount over the continuation induced
at $h$ by following $\widetilde{\alpha}$. Therefore, $\widetilde{\alpha}$
is an empty-threat-free computational Nash equilibrium of $\GammaPK$.
\end{proof}

\end{document}